\documentclass[twocolumn,nofootinbib,showpacs,preprintnumbers,amsmath,amssymb]{revtex4}

\usepackage{graphicx}
\usepackage{dcolumn}
\usepackage{bm}
\usepackage{amssymb}
\usepackage{amsmath}

\def\bew{\begin{widetext}}    
\def\eew{\end{widetext}}      

\renewcommand{\vec}[1]{{\bf #1}}       
\def\beq{\begin{eqnarray}}    
\def\eeq{\end{eqnarray}}      

\newcommand{\OM}{\Omega_M}
\newcommand{\Om}{\Omega_m}

\newcommand{\OMo}{\Omega_{M}^0}

\newcommand{\OLo}{\Omega_{\Lambda}^0}
\newcommand{\OX}{\Omega_{X}}
\newcommand{\OXo}{\Omega_{X}^0}

\newcommand{\OD}{\Omega_{D}}
\newcommand{\ODo}{\Omega_{D}^0}

\newcommand{\rc}{\rho_c}
\newcommand{\rco}{\rho_{c}^0}

\newcommand{\rM}{\rho_M}

\newcommand{\rD}{\rho_D}

\newcommand{\rX}{\rho_X}

\newcommand{\wX}{\omega_X}
\newcommand{\wm}{\omega_m}

\newcommand{\rL}{\rho_{\CC}}

\newcommand{\rLo}{\rho_{\CC}^0}
\newcommand{\pD}{p_D}
\newcommand{\wD}{\omega_D}

\newcommand{\wL}{\omega_{\CC}}
\newcommand{\CC}{\Lambda}

\newcommand{\we}{w_e}

\newcommand{\tOM}{\tilde{\Omega}_M}

\newcommand{\tOL}{\tilde{\Omega}_{\CC}}
\newcommand{\tOD}{\tilde{\Omega}_{D}}

\def\de{\delta}

\def\La{\Lambda}

\def\Ga{\Gamma}

\def\La{\Lambda}

\def\Om{\Omega}

\def\na{\nabla}

\def\hh{\hat{h}}

\def\csd{c_{s}^2}
\def\cad{c_{a}^2}
\def\cs{c_{s}^2}

\def\dt{\stackrel{\circ}}

\begin{document}




\hyphenation{cos-mo-lo-gi-cal
sig-ni-fi-cant quin-tes-sen-ce}









\title{Dark energy perturbations and cosmic coincidence}

\author{Javier Grande}
 \email{jgrande@ecm.ub.es}
\author{Ana Pelinson}%
 \email{apelinson@ecm.ub.es}
\author{Joan Sol\`{a}}
 \email{sola@ecm.ub.es}
\affiliation{%
High Energy Physics Group, Dept. ECM, and Institut de Ci{\`e}ncies del Cosmos\\
Univ. de Barcelona, Av. Diagonal 647, E-08028 Barcelona, Catalonia, Spain}%

\date{\today}

\begin{abstract}
While there is
plentiful evidence in all fronts of experimental cosmology for the
existence of a non-vanishing dark energy (DE) density $\rD$ in the
Universe, we are still far away from having a fundamental
understanding of its ultimate nature and of its current value, not
even of the puzzling fact that $\rD$ is so close to the matter
energy density $\rM$ at the present time (i.e. the so-called
``cosmic coincidence'' problem). The resolution of some of these
cosmic conundrums suggests that the DE must have some (mild)
dynamical behavior at the present time. In this paper, we examine
some general properties of the simultaneous set of matter and DE
perturbations $(\delta\rho_M, \delta\rho_D)$ for a multicomponent DE
fluid. Next we put these properties to the test within the context
of a non-trivial model of dynamical DE (the $\CC$XCDM model) which
has been previously studied in the literature. By requiring that the
coupled system of perturbation equations for $\delta\rM$ and
$\delta\rD$ has a smooth solution throughout the entire cosmological
evolution, that the matter power spectrum is consistent with the
data on structure formation and that the ``coincidence ratio''
$r=\rD/\rM$ stays bounded and not unnaturally high, we are able to
determine a well-defined region of the parameter space where the
model can solve the cosmic coincidence problem in full compatibility
with all known cosmological data.
\end{abstract}

\maketitle


\vskip 6mm

 \noindent \section{Introduction}
 \label{Introduction}

Undoubtedly the most prominent accomplishment of modern cosmology to
date has been to provide strong indirect support for the existence
of both dark matter (DM) and dark energy (DE) from independent data
sets derived from the observation of distant
supernovae\,\cite{Supernovae}, the anisotropies of the
CMB\,\cite{WMAP3}, the lensing effects on the propagation of light
through weak gravitational fields\,\cite{Lensing}, and the inventory
of cosmic matter from the large scale structures (LSS) of the
Universe\,\cite{Cole05,Lahav02}. But, in spite of these outstanding
achievements, modern cosmology still fails to understand the
ultimate physical nature of the components that build up the
mysterious dark side of the Universe, most conspicuously the DE
component of which the first significant experimental evidence was
reported $10$ years ago from supernovae observations. The current
estimates of the DE energy density yield $\rD^{\rm exp}\simeq
(2.4\times 10^{-3}\,eV)^4$ and it is believed that it constitutes
roughly $70\%$ of the total energy density budget for an essentially
flat Universe. The big question now is: what is it from the point of
view of fundamental physics? One possibility is that it is the
ground state energy density associated to the quantum field theory
(QFT) vacuum and, in this case, it is traditional to associate $\rD$
to $\rL=\CC/8\pi\,G$, where $\CC$ is the cosmological constant (CC)
term in Einstein's equations. The problem, however, is that the
typical value of the (renormalized) vacuum energy in all known
realistic QFT's is much bigger than the experimental value. For
example, the energy density associated to the Higgs potential of the
Standard Model (SM) of electroweak interactions is more than fifty
orders of magnitude larger than the measured value of $\rD$.

Another generic proposal (with many ramifications) is the
possibility that the DE stands for the current value of the energy
density of some slowly evolving, homogeneous and isotropic scalar
field (or collection of them). Scalar fields appeared first as
dynamical adjustment mechanisms for the CC\,\cite{Dolgov,PSW} and
later gave rise to the notion of quintessence\,\cite{quintessence}.
While this idea has its own merits (specially concerning the
dynamical character that confers to the DE) it has also its own
drawbacks. The most obvious one (often completely ignored) is that
the vacuum energy of the SM is still there and, therefore, the
quintessence field just adds up more trouble to the whole
fine-tuning CC problem\,\cite{weinRMP,CCRev}!

Next-to-leading is the ``cosmological coincidence problem'', or the
problem of understanding why the presently measured value of the DE
is so close to the matter density. One expects that this problem can
be alleviated by assuming that $\rD$ is actually a dynamical
quantity. While quintessence is the traditionally explored option,
in this paper we entertain the possibility that such dynamics could
be the result of the so-called cosmological ``constants'' (like
$\CC$, $G$,...) being actually variable. It has been proven in
\,\cite{SS12} that this possibility can perfectly mimic
quintessence. It means that we stay with the $\CC$ parameter and
make it ``running'', for example through quantum
effects\,\cite{cosm,JHEPCC1,Babic}\,\footnote{ For a general
discussion, see \,\cite{SSIRGAC06,ShS08}.}. However, in
\,\cite{LXCDM12} it was shown that, in order to have an impact on
the coincidence problem, the total DE in this context should be
conceived as a composite fluid made out of a running $\CC$ and
another entity $X$, with some effective equation of state (EOS)
parameter $\wX$, such that the total DE density and pressure read
$\rD=\rL+\rX$ and $\pD=-\rL+\wX\rX$, respectively. We call this
system the $\CC$XCDM model\,\cite{LXCDM12}. Let us emphasize that
$X$ (called ``the cosmon'') is not necessarily a fundamental entity;
in particular, it need not be an elementary scalar field. As
remarked in \,\cite{LXCDM12}, $X$ could represent the effective
behavior of higher order terms in the effective action (including
non-local ones). This is conceivable, since the Bianchi identity
enforces a relation between all dynamical components that enter the
effective structure of the energy-momentum tensor in Einstein's
equations, in particular between the evolving $\CC$ and other terms
that could emerge after we embed General Relativity in a more
general framework\,\cite{Gruni,fossil}. Therefore, at this level, we
do not impose a microscopic description for $X$ and in this way the
treatment becomes more general\,\footnote{See e.g.
\cite{Ratra08ab,Percival07,Zhang08a} for recent constraints on
$\CC$CDM, XCDM and quintessence-like models. The margin for the
energy densities and EOS parameter $\wX$ is still quite high. In the
$\CC$XCDM case, the fact that $\CC$ is running provides an even
wider range of phenomenological possibilities.}. The only condition
defining $X$ is the DE conservation law, namely we assume that
$\rD=\rL+\rX$ is the covariantly self-conserved total DE density.

In this paper, we analyze the combined dynamics of DE and matter
density perturbations for such conserved DE density $\rD$. The
present study goes beyond the approximate treatment presented in
\cite{GOPS}, where we neglected the DE perturbations and estimated
the matter perturbations of the $\CC$XCDM model using an effective
(variable) EOS $\we$ for the composite fluid $(\rL,\rX)$. The main
result was that a sizeable portion of parameter space was still
compatible with a possible solution of the cosmic coincidence
problem. The ``effective approach'' that we employed in \cite{GOPS}
was based on three essential ingredients: i) the use of the
effective EOS representation of cosmologies with variable
cosmological parameters\,\cite{SS12}; ii) the calculation of the
growth of matter density fluctuations using the effective EOS of the
DE\,\cite{LinderJenkins03}; and iii) the application of the,
so-called, ``F-test'' to compare the model with the LSS data, i.e.
the condition that the linear bias parameter, $b^2(z)=
P_{GG}/P_{MM}$ does not deviate from the $\CC$CDM value by more than
$10\%$ at $z=0$, where $P_{MM}\propto (\delta\rho_M/\rho_M)^2$ is
the matter power spectrum and $P_{GG}$ is the galaxy fluctuation
power spectrum\,\cite{Cole05,Lahav02} -- see \cite{GOPS,Ftest} for
details. This three-step methodology turned out to be an efficient
streamlined strategy to further constrain the region of the original
parameter space\,\cite{LXCDM12}. However, there remained to perform
a full fledged analysis of the system of cosmological perturbations
in which the DE and matter fluctuations are coupled in a dynamical
way. This kind of analysis is presented here.

The structure of the paper is as follows. In the next section, we
outline the meaning of the cosmic coincidence problem within the
general setting of the cosmological constant problem. In section
\ref{sect:perturbations}, the basic equations for cosmological
perturbations of a multicomponent fluid in the linear regime are
introduced. In section \ref{sect:effEOS}, we describe the general
framework for addressing cosmological perturbations of a composite
DE fluid with an effective equation of state (EOS). In section
\ref{sect:genericfeatures}, we describe some generic features of the
cosmological perturbations for the dark energy component. The
particular setup of the $\CC$XCDM model is focused in section
\ref{sect:LXCDM}. In sections \ref{sect:perturbLXCDM} and
\ref{numerical}, we put the $\CC$XCDM model to the stringent test of
cosmological perturbations and show that the corresponding region of
parameter space becomes further reduced. Most important, in this
region the model is compatible with all known observational data
and, therefore, the $\CC$XCDM proposal can be finally presented as a
robust candidate model for solving the cosmic coincidence problem.
In section \ref{DEfluct}, we offer a deeper insight into the
correlation of matter and DE perturbations. In the last section, we
present the final discussion and deliver our conclusions.

\noindent \section{The coincidence problem as a part of the big CC
problem} \label{sect:cosmiccoincicence}

The cosmic coincidence problem is a riddle, wrapped in the
polyhedric mystery of the Cosmological Constant
Problem\,\cite{weinRMP,CCRev}, which has many faces. Indeed, we
should clearly distinguish between the two main aspects which are
hidden in the cosmological constant (CC) problem. In the first
place, we have the ``old CC problem'' (the ugliest face of the CC
conundrum!), i.e. the formidable task of trying to explain the
relatively small (for Particle Physics standards) measured value of
$\rL$ or, more generally, of the DE density\,\cite{Supernovae},
roughly $\rD^{\rm exp}\sim 10^{-47}\,GeV^4$, after the many phase
transitions that our Universe has undergone since the very early
times, in particular the electroweak Higgs phase transition
associated to the Standard Model of Particle Physics, whose natural
value is in the ballpark of $\rho_{EW}\sim G_F^{-2}\sim 10^9\,GeV^4$
($G_F$ being Fermi's constant). The discrepancy $\rho_{EW}/\rD^{\rm
exp}$, which amounts to some $56$ orders of magnitude, is the
biggest enigma of fundamental physics ever!\,\footnote{See e.g.
\cite{JHEPCC1,SSIRGAC06,ShS08} for a summarized presentation with
specific insights closely related to the present approach.} Apart
from the induced CC contribution from phase transitions, we have the
pure vacuum-to-vacuum quantum effects. Since the (renormalized) zero
point energy of a free particle of mass $m$ contributes $\sim m^4$
to the vacuum energy density\,\cite{cosm,JHEPCC1}, it turns out that
even a free electron contributes an amount more than thirty orders
of magnitude larger than the aforementioned experimental value of
$\rD$. Only light neutrinos $m_{\nu}\sim 10^{-3}\,eV$, or scalar
particles of similar mass, could contribute just the right amount,
namely if these particles would be the sole active degrees of
freedom in our present cold Universe (see\,\cite{cosm}).

On the other hand, the cosmic coincidence problem\,\cite{Steinhardt}
is that second (``minor'') aspect of the CC problem addressing the
specific question: ``why just now?'', i.e.  why do we find ourselves
in an epoch $t=t_0$ where the DE density is similar to the matter
density, $\rD(t_0)\simeq\rM(t_0)$? In view of the rapidly decreasing
value of $\rM(a)\sim 1/a^3$, where $a=a(t)$ is the scale factor, it
is quite puzzling to observe that its current value is precisely of
the same order of magnitude as the vacuum energy or, in general, the
dark energy density $\rD$. It is convenient to define the ``cosmic
coincidence ratio''
\begin{equation}\label{ra}
r(a)=\frac{\rD(a)}{\rM(a)}=\frac{\Omega_D(a)}{\Omega_M(a)}\,,
\end{equation}
where $(\Omega_D(a), \Omega_M(a))$ are the corresponding densities
normalized with respect to the current critical density
$\rho^0_c\equiv 3H^2_0/8\pi G$. For $\OMo\simeq 0.3$ and $\ODo\simeq
0.7$, we have $r_0\simeq 2.3$, which is of ${\cal O}(1)$. However,
in the standard cosmological $\CC$CDM model, where $\OD$ is constant
and $\OM(a\to\infty)\rightarrow 0$, the ratio $r$ grows unboundedly
with the expansion of the Universe. So the fact that $r_0={\cal
O}(1)$ is regarded as a puzzle because it suggests that $t=t_0$ is a
very special epoch of our Universe. One could also consider the
inverse ratio $r^{-1}={\rM(a)}/{\rD(a)}$, which goes to zero with
the expansion. The coincidence problem can be equivalently
formulated either by asking why $r$ is not very large now or why
$r^{-1}$ is not very small. Solving the coincidence problem would be
to find either 1) a concrete explanation for $r$ and $r^{-1}$ being
of order one at present within the standard cosmological model, or
2) a modified cosmological model (compatible with all known
cosmological data) insuring that these ratios do not undergo a
substantial change, say by more than one order of magnitude or so,
for a very long period of the cosmic history that includes our time.

In a very simplified way, let us summarize some of the possible
avenues that have been entertained to cope with the coincidence
puzzle:

\begin{itemize}

\item Quintessence and the like \,\cite{quintessence,TrackingQ,
InteractingQ,Oscillatory,kessence}. One postulates the existence of
a set of cosmological scalar fields $\phi_{i}$ essentially unrelated
to the rest of the particle physics world. The DE produced by these
fields has an effective EOS parameter $\wD\gtrsim -1$ which causes
$\rD$ to decrease always with the expansion (i.e. $d\rho_D/da<0$),
but at a pace slower (on average) than that of the background
matter. Thus, it finally catches up with it and $\rD$ emerges to
surface, i.e. the condition $\rD>\rM$ eventually holds (presumably
near our time). In this framework, there is the possibility of
self-adjusting and tracker solutions\,\cite{quintessence,TrackingQ},
where the DE keeps track of the matter behavior and ultimately
dominates the Universe. It requires to take some special forms of
the potential, and in some cases the Lagrangian involves
non-canonical kinetic terms. For example, in the simple case of a
single scalar field and the exponential potential $V(\phi)\sim
\exp{\left(-\lambda\phi/M_P\right)}$ one finds that the coincidence
ratio becomes fixed at the value
\begin{equation}\label{rexp}
r=\frac{3(1+\wm)}{\lambda^2-3(1+\wm)}\,,
\end{equation}
where $\wm$ is the EOS of the background matter (i.e. $0$ or $1/3$,
depending on whether cold or relativistic matter dominates,
respectively). So, at the present time, $r=3/(\lambda^2-3)$, and by
appropriate choice of $\lambda$ one can match the current
experimental value. But of course the choice of the potential was
rather peculiar and the field $\phi$ itself is completely \textit{ad
hoc}. Moreover, it has a mass $m_{\phi}=\sqrt{V''(\phi)}\sim H\sim
10^{-33}\,eV$ (as it follows from a self-consistent solution of
Einstein's equations); such mass scale is $30$ orders of magnitude
below the mass scale associated to the DE, which is $\rD^{1/4}\sim
10^{-3}\,eV$. In this sense, it looks a bit unnatural to aim at
solving the CC problem by introducing a field whose extremely tiny
mass creates another cosmological puzzle.

On the other hand, within the context of interactive quintessence
models\,\cite{InteractingQ} (whose main leitmotif is precisely
trying to cure the coincidence problem), the coupling of $\phi_i$
and the matter components makes allowance for energy exchange
between the two kinds of fields, and as a result the ratio
(\ref{ra}) can be constant or slowly variable, whereas in other
implementations one can achieve an oscillatory tracking behavior of
$r$, although the construction is essentially \textit{ad
hoc}\,\cite{Oscillatory}. Another generalization leads to k-essence
models\,\cite{kessence} (characterized by non-canonical kinetic
terms), where fine tuning problems in the tracking can be disposed
of but the dominant background component can be tracked only up to
matter-radiation equality and is lost immediately afterwards (as the
DE is immediately prompted into a CC-like behavior). In one way or
another, however, all variants of quintessence suffer from several
drawbacks, and in particular from the following generic one: they
assume (somehow implicitly) that the remaining fields of the
particle physics spectrum (i.e. those which were already there from
the very beginning) have nothing to do with the CC problem. As a
result of such a bold assumption, the (likely real) vacuum problem
of the conventional fields in QFT is merely traded for the (likely
fictitious) vacuum problem of quintessence, {which is no less acute
because no real explanation is provided for the smallness of the
current $\rD$ value versus $m^4$ (where $m$ is any typical mass
scale in Particle Physics). Hence we are back to the same kind of CC
problem we started with}.

\item Phantom energy\,\cite{Phantom}. It is motivated by the fact
that, observationally speaking, the effective EOS of the DE cannot
be excluded to satisfy $\wD\lesssim -1$ near our present time. As
indicated above, many quintessence-like models in reality are hybrid
constructions containing a mixture of fields with a phantom
component. The reason is that one wants to give allowance for a
``CC-crossing'' $\ \wD=-1$ near our time. While phantom energy
shares with quintessence the use of scalar fields $\phi_i$, here the
DE produced by these fields is always \textit{increasing} with the
expansion, $d\rho_D/da>0$, even after the relation $\rD>\rM$ is
fulfilled. The consequence of this ever growing behavior of the dark
component is that one ends up with a superaccelerated expansion of
the Universe that triggers an eventual disruption of all forms of
matter (the so-called ``Big Rip''). When computing the fraction of
the lifetime of the Universe where the ratio (\ref{ra}) stays within
given bounds before the ``doomsday'', one finds that it can be
sizeable.

\item Non-local theories. There is some renewed interest in this
kind of theories in which the emphasis is placed on the existence of
possible non-local structures in the effective action
\,\cite{DeserWoodard}. It has recently been emphasized in
\cite{ShS08} that the dynamical evolution of the vacuum energy
should come from a resummation of terms in the effective action
leading to non-local contributions of the form $R\, {\cal F}({\cal
G}_0\, R)$, for some unknown function ${\cal F}$ of dimension $2$,
where ${\cal G}_0$ is the massless Green's function (${\cal G}_0\sim
1/\Box$). The canonical possibility would be ${\cal F}= M^2 {\cal
G}_0 R$, where $M$ is a parameter with dimension of mass. This
situation leads to an effective evolution of the CC of the form
$\Delta\rL\sim M^2\, H^2$ during the matter dominated epoch, whereas
in the radiation era the effective CC would approximately be zero
(because $R\sim T^{\mu}_{\mu}\simeq 0$ for relativistic matter, see
(\ref{EinsteinEq2})-(\ref{TraceEMT}) below). As a result, the
coincidence puzzle could somehow be understood from the fact that
the CC may start to be preponderant at some point once the onset of
the matter dominated epoch is left behind.

\item Of course many other ideas have been explored. For instance, one may
introduce special fluids with very peculiar EOS, such as the
Chaplygin gas\,\cite{Chaplygin}, which behaves as pressureless
matter at early times ($\wD\simeq 0$) and as vacuum energy at
present ($\wD\simeq -1$). Although there is some connection with
braneworld cosmology, this proposal suffers from the same problem as
quintessence in that it supersedes the vacuum state of traditional
fields by the new vacuum of that peculiar fluid. Finally, let us
mention the Anthropic models, which fall in a quite different
category, in the sense that one does not look for a solution of the
coincidence problem exclusively from first principles of QFT or
string theory, but rather through the interplay of the ``human
factor''. Basically, one ties the value of the ratio (\ref{ra}) to
the time when the conditions arise for the development of
intelligent life in the Universe, in particular of cosmologists
making observations of the cosmos. This variant has also a long
story, but we shall refrain from entering the details, see e.g.
\,\cite{weinRMP,Anthropic}.

\end{itemize}

\noindent \section{Cosmological perturbations for a multicomponent
fluid in the linear regime} \label{sect:perturbations}

In the remaining of the paper we concentrate on studying some
general properties of the cosmological perturbations, both of matter
and DE, and the implications they may have on the coincidence
problem within models characterized by having running vacuum energy
and other DE components. According to cosmological perturbation
theory, all energy density components, including the dark energy,
should fluctuate and contribute to the growth of the large-scale
cosmological structures. In this section, we discuss the general
framework of linear density perturbations in models composed of a
multicomponent DE fluid besides the canonical matter.

In the following we use the standard metric perturbation approach
\cite{cosmobooks} and consider simultaneous density and pressure
perturbations
\beq \label{dppert} \rho_N\to\,\rho_N+\de\rho_N \,, \qquad
p_N\to\,p_N+\de p_N \,, \eeq
for all the components ($N=1,2,...$) of the  fluid, including matter
and all possible contributions from the multicomponent DE part. At
the same time, we have metric and velocity perturbations for each
component:
\beq g^B_{\mu\nu}\to\, g_{\mu\nu}=g^B_{\mu\nu}+\delta g_{\mu\nu}\,,
\qquad U_N^{\mu}\to U_N^{\mu}+\de U_N^{\mu}\,, \label{mfpert} \eeq
where $g^B_{\mu\nu}$ is the background metric. The 4-vector velocity
$U_N^{\mu}$ in the comoving coordinates has {the following
components and perturbations $(U_N^0,U_N^i)=(1,0)\to \de
U_N^{\mu}=(0,v_N^i)$, where} $v_N^i$ is the three-velocity of the
$N$th component of the fluid in the chosen coordinate system. As a
background space-time, we assume the homogeneous and isotropic
Friedmann-Lema\^itre-Robertson-Walker (FLRW) metric with flat space
section, hence $g^B_{\mu\nu}=\mbox{diag}\,
\big\{1,\,-a^2(t)\,\de_{ij}\big\}$, where $a$ is the scale factor.

In order to derive the set of perturbed equations, let us first
introduce Einstein's equations:
\beq \label{EinsteinEq} R_{\mu\nu}-\frac{1}{2}R\,g_{\mu\nu}=8\pi G\,
{T}_{\mu\nu}\,,
 \eeq
where $G$ is the Newton constant and ${T}_{\mu\nu}$ is the total
energy-momentum tensor of matter and dark energy. Both the
background and perturbed metric are  assumed to satisfy these
equations. The total energy-momentum tensor of the system is assumed
to be the sum of the perfect fluid form for each component:
\begin{equation} \label{EMT} {T}_{\mu\nu} = \sum_N {T}^N_{\mu\nu}=\sum_N\left[ -
p_N\,g_{\mu\nu}+\big(\rho_N +
p_N\,\big)\,U_{\mu}^N\,U_{\nu}^N\right]\,. \end{equation}
The components of $T^{\mu}_{\nu}$ are the following:
\beq {T}^0_{0}&=&\sum_N \rho_N \equiv \rho_T \,,\\
{T}^i_j&=&-\sum_N p_N \,\de^i_j\equiv -p_T\,\de^i_j \,, \eeq
where $ \rho_T$ and $p_T$ are the total energy density and pressure,
respectively.

Perturbations on the metric and on the energy-momentum tensor are
uniquely defined for a given perturbed space-time, provided we make
a gauge choice. The latter means that we choose a specific
coordinate system; in this way, four out of the $10$ components
$\delta g_{\mu\nu}\equiv h_{\mu\nu}$ of the metric perturbation can
be fixed at will. Here we have adopted the synchronous gauge, widely
used in the literature, in which the four preassigned values of the
metric perturbations are $h_{0i}=0$ and $h_{0 0}=0$. Setting
$h_{\mu\nu}=-a^2\,\chi_{\mu\nu}$, in this gauge the perturbed,
spatially flat, FLRW metric takes on the form
\beq\label{FLRWsynch} ds^2 & = &g_{\mu\nu}\,dx^{\mu}dx^{\nu}=dt^2
-a^2(t)\left(\delta_{ij}+\chi_{ij}\right)dx^{i}dx^{j}
\nonumber\\
& = &
a^2(\eta)\left[d\eta^2-\left(\delta_{ij}+\chi_{ij}\right)dx^{i}dx^{j}\right]\,,
\eeq
where in the last equality we have expressed the result also in
terms of the conformal time $\eta$, defined through $d\eta=dt/a$. We
may compare (\ref{FLRWsynch}) with the most general perturbation of
the spatially flat FLRW metric, \beq\label{FLRWpert}
ds^2  &=&a^2(\eta)[(1+2\psi)d\eta^2-\omega_{i}\,d\eta\,dx^i\nonumber\\
&& -\, \left((1-2\phi)\delta_{ij}+\chi_{ij}\right)dx^{i}dx^{j}]\,,
\eeq consisting of the $10$ degrees of freedom associated to the two
scalar functions $\psi$, $\phi$, the three components of the vector
function $\omega_i\, (i=1,2,3)$, and the five components of the
traceless $\chi_{ij}$. Clearly, the synchronous gauge
(\ref{FLRWsynch}) is obtained by setting $\psi=0$, $\omega_i=0$ and
absorbing the function $\phi$ into the trace of $\chi_{ij}$. In this
way, $\chi_{ij}$ in (\ref{FLRWsynch}) contributes six degrees of
freedom. As we will see in a moment, in practice only the
nonvanishing trace of the metric disturbance will be necessary to
perform the analysis of cosmic perturbations in this gauge. To
within first order of perturbation theory, such a trace is given by
\begin{equation}\label{trace}
 h\equiv g^{\mu\nu}\,h_{\mu\nu}=
 g^{ij}\,h_{ij}=-\frac{h_{ii}}{a^2}=\chi_{ii}\,,
\end{equation}
where $g^{ij}$ is the inverse of
$g_{ij}=-a^2(t)\,(\delta_{ij}+\chi_{ij})$, and it is understood that
the repeated Latin indices are summed over $1,2,3$.

For the physical interpretation, notice that the synchronous gauge
is associated to a coordinate system in which the cosmic time
coordinate is comoving with the fluid particles ($g_{00}=1$, i.e.
$\psi=0$), which is the reason for its name and also explains why
this gauge does not have an obvious Newtonian limit. In fact, this
gauge choice is generally appropriate for the study of fluctuations
whose wavelength is larger than the Hubble radius ($\lambda\gg
d_H\equiv H^{-1}$). Actually, any mode satisfies this condition at
sufficiently early epochs, and in this regime the effects of the
space-time curvature are unavoidable.

Next we wish to compute the perturbations in the synchronous gauge.
To start with, it is convenient to rewrite Einstein's equations
(\ref{EinsteinEq}) as follows,
\beq \label{EinsteinEq2} R_{\mu\nu}=8\pi G\, {S}_{\mu\nu}\,,\ \ \ \
{S}_{\mu\nu}\equiv {T}_{\mu\nu}-\frac{1}{2}\,g_{\mu\nu}\,T\,,
 \eeq
where $T=T^{\mu}_{\mu}$ is the trace of (\ref{EMT}), hence
\beq \label{TraceEMT} {T} = \sum_N\left(\rho_N-3\,p_N\right)\,. \eeq
For the calculation of the perturbations, we can use any of the
components of Einstein's equations. However, since we are going to
use the conservation law $\na_\mu {T}^\mu_\nu = 0$ to derive
additional fluctuation equations, it is convenient to perturb the
$(00)$-component of (\ref{EinsteinEq2}) because the other components
are well-known not to be independent of the conservation law. Thus,
using (\ref{EMT}), (\ref{EinsteinEq2}) and (\ref{TraceEMT}) we
obtain
\beq\label{Szz}
S_{00} &=& T_{00}-\frac12\,T=\frac12\,\sum_N\left(\rho_N+3\,p_N\right)\nonumber\\
&\Rightarrow& \delta
S_{00}=\frac12\,\sum_N\left(\delta\rho_N+3\delta\,p_N\right)\,. \eeq
On the other hand, a straightforward calculation shows that the
perturbation of the $(00)$-component of the Ricci tensor can be
written as follows:
\begin{equation}\label{dRmn}
\delta R_{00}=-\frac12\,\frac{\partial^2 h}{\partial
t^2}-H\,\frac{\partial h}{\partial t}=\frac12\,\dt{\hh}+ H\,\hh\,,
\end{equation}
where we have used (\ref{trace}) and defined the ``hat variable''
\beq \label{hdef} \hh \equiv  \frac{\partial}{\partial
t}\left(\frac{h_{ii}}{a^2}\right)=-\dt{h}\,. \eeq
The overhead circle ( $^{\circ}$ ) indicates partial differentiation
with respect to the cosmic time (i.e. $\dt{f}\equiv \partial
f/\partial t$, for any $f$), in order to distinguish it from other
differentiations to be used later on. Therefore, $H =\ \dt{a}/a$ is
the ordinary expansion rate in the cosmic time $t$.

Since the fluctuations  $\delta S_{00}$ and $\delta R_{00}$ from
(\ref{Szz}) and (\ref{dRmn}) are constrained by (\ref{EinsteinEq2}),
we find
\beq \label{PertEeq} \dt{\hh} + 2 H \hh = 8\pi G\, \sum_N (\de
\rho_{N}+3 \de p_N) \,. \eeq
If we substitute (\ref{hdef}) in the previous expression, a second
order differential equation in the original variable (\ref{trace})
ensues. In terms of the conformal time $\eta$, it can be written as
follows:
\beq \label{PertEeqConf} \ddot{h} +  {\cal H} \dot{h} = - 8\pi
G\,a^2\, \sum_N (\de \rho_{N}+3 \de p_N) \,, \eeq
where the dot ( $\dot{}$ ) indicates differentiation with respect to
$\eta$ (i.e. $\dot{f}\equiv df/d\eta$) and ${\cal H}\equiv
\dot{a}/a=H\,a$ is the expansion rate in the conformal time.

The Friedmann equation can be written in terms of the normalized
densities as
\beq {H^2} = \frac{8\pi G}{3}\rho_{T} = {H_0^2}\, \sum_N \Omega_{N}
\,,\label{friedeq} \eeq
where $H_0$ is the present value of the Hubble parameter and
$\Omega_{N}\equiv \rho_{N}/\rho^0_c$ are the normalized densities
with respect to the current critical density $\rho^0_c\equiv
3H^2_0/8\pi G$.

The subsequent step is to perform perturbations on the conservation
law $\na_\mu {T}^{\mu\nu} = 0$, as this will provide the additional
independent equations. Using (\ref{EMT}), the previous law reads
explicitly as follows:
\beq
\label{EnergyCons2}
&&\sum_N\Big\{\nabla_{\mu}(\rho_N+p_N)\,U_N^{\mu}U_N^{\nu}+(\rho_N+p_N)\nonumber\\
&&\phantom{\sum_N\Big\{}\times
\left[U_N^{\nu}\na_\mu
{U}^\mu_N+U_N^{\mu}\na_\mu
{U}^\nu_N\right]\Big\}=g^{\mu\nu}\,\sum_N\,\na_{\mu}\,p_N\,.\nonumber\\
\eeq
For any four-velocity vector, we have $U^{\mu}_N\,U_{\mu}^N=1$ and,
therefore, we have the orthogonality relation
$U_{\nu}^N\na_{\mu}U^{\nu}_N=0$. In this way, by contracting
Eq.\,(\ref{EnergyCons2}) with $U_{\nu}^N$ we find the simpler result
\begin{equation}\label{conserv3}
\sum_N\left[U_N^\mu\,\na_{\mu}\,\rho_N+(\rho_N+p_N)\na_\mu
{U}^\mu_N\right]=0\,.
\end{equation}
Let us emphasize that the sum $\sum_N$ in this equation need not run
necessarily over all the terms of the cosmic fluid. It may hold for
particular subsets of fluid components that are overall
self-conserved. In particular, it could even hold for each
component, if they would be individually conserved. In our case, it
applies to the specific matter component and also, collectively, to
the multicomponent DE part.  It is straightforward to check that, in
the FLRW metric, we have
\begin{equation}\label{nablaU3H}
\na_\mu {U}^\mu_N=3\,H\,\ \ \ (N=1,2,...)\,.
\end{equation}
Using this relation, it is immediate to see that, in the co-moving
frame, Eq.(\ref{conserv3}) boils down to
\beq\label{EnergyCons} \sum_N\left[ \dt{\rho}_N + 3 H
(\rho_N+p_N)\right] =0\,. \eeq
Moreover, perturbing (\ref{nablaU3H}) in the synchronous gauge, we
find (using $\delta\Gamma^{\mu}_{\mu\,0}=-\hh/2$ for the perturbed
Christoffel symbol involved in the covariant derivative) the useful
result
\begin{equation}\label{pertH}
\delta H=\frac13\,\delta\left(\na_\mu
{U}^\mu_N\right)=\frac13\,\left(\theta_N-\frac{\hh}{2}\right)\,,
\end{equation}
where we have introduced the notation ${\theta}_N\equiv \,\nabla_\mu
(\de U_N^\mu)=\nabla_i (\de U_N^i) \,$ (with $\delta U_N^0=0$) for
the covariant derivative of the perturbed three-velocity $\de
U_N^i=v_N^i$. Equipped with these formulas, the perturbed
Eq.\,(\ref{EnergyCons}) immediately leads to
\begin{equation} \label{EConsPert1}
\sum_N\left[\dt{\de\rho_N}+(\rho_N+p_N)\left({\theta}_N-\frac{\hh}{2}\right)
+3H(\de\rho_N+\de p_N)\right] =0\,.\end{equation}
The previous result could have equivalently been obtained by setting
$\nu=0$ in (\ref{EnergyCons2}) and perturbing the corresponding
equation. An independent relation can be obtained by setting $\nu=i$
in (\ref{EnergyCons2}) within the co-moving frame and carrying out
the perturbation. Using the relevant Christoffel symbol
$\Gamma^{i}_{0 j}= H\,\delta_{ij}$ and Eq.(\ref{nablaU3H}) we
obtain, after some calculations,
\beq\label{mueqi}&&\sum_N\bigg\{(\rho_N+p_N)\dt{\delta U^i}_N+\Big[\dt{\rho}_N +\dt{p}_N+5H(\rho_N+p_N)\Big]\nonumber\\
&&\phantom{\sum_N\bigg\{}\times\delta{U^i_N}\bigg\}=\sum_N \na^i\delta p_N\,.\eeq
The final step is obtained by computing the divergence $\na_i$ on
both sides of this equation. To within first order of perturbation
theory, we obtain $\na_i\na^i\delta p_N=g^{ik}\na_i\na_k\,\delta
p_N=-(1/a^2)\na^2\delta p_N$ for the action of $\na_i$ on the
\textit{r.h.s.} of (\ref{mueqi}), whereas on its \textit{l.h.s.} we
can use the variable $\theta_N$ previously defined. Moreover, we
have $\nabla_i\dt{\delta U^i}_N\, =\,\dt{\theta}_N$ owing to the
commutativity of coordinate differentiation and perturbation
operations. In this way, the final outcome reads
\beq \label{EConsPert2}
&&\sum_N\bigg\{(\rho_N+p_N)\dt{\theta}_N+\Big[\dt{\rho}_N+\dt{p}_N+5H
(\rho_N+p_N) \Big]\nonumber\\
&&\phantom{\sum_N\bigg\{}\times{\theta}_N\bigg\}=\frac{k^2}{a^2}\sum_N\de
p_N,
\eeq
where it is furthermore understood that we have used the Fourier
decomposition for all the perturbation variables:
\beq\label{planewave} \de f(t,\vec{x})=\int \frac{d^3
k}{{(2\pi)}^3}\ \de {f(t,k)}\,e^{i \vec{k}\cdot \vec{x}} \,, \eeq
where $\delta f=(\hh,\delta\rho_N, \delta p_N, \theta_N$).
In Fourier space, the perturbation variables are denoted with the
same notation, but they are the Fourier transforms of the original
ones, so their arguments are $t$ and $k$ because the space variable
$\vec{x}$ has been traded for the wave number $k\equiv|\vec{k}|$.
The latter will be measured in units of $h\,$Mpc$^{-1}$, where
$h\simeq 0.7$ is the reduced Hubble constant -- not to be confused
with the trace of the synchronous perturbed metric,
Eq.\,(\ref{trace}). In these units, the linear regime corresponds to
length scales $\ell\sim k^{-1}$ with wave numbers $k
<0.2\,h\,$Mpc$^{-1}$, i.e. $\ell>5\,h^{-1}\,Mpc$. Notice that, if
desired, one can easily rewrite the above perturbation equations
(\ref{EConsPert1}) and (\ref{EConsPert2}) in conformal time simply
by using $\dt{f}=\dot{f}/a$ for any $f$.


We have obtained three basic sets of perturbation equations
(\ref{PertEeq}), (\ref{EConsPert1}) and (\ref{EConsPert2}) for the
four kinds of perturbation variables $(\hh(t,k),\
\delta\rho_N(t,k),\ \delta p_N(t,k),\ \theta_N(t,k))$. It is thus
clear that the evolution of the cosmic perturbations can be
completely specified only after we assume some relation of the
pressure perturbation $\de p_N$ and the density perturbation $\de
\rho_N$ for each fluid. If the perturbations are adiabatic, then
that relation is simply
\begin{equation}
\de p_N=c_{a,N}^2\de \rho_N\,,
\end{equation}
where $c_{a,N}^2$ is the adiabatic speed of sound for each fluid,
defined as:
\beq\label{adspeed} c_{a,N}^2\equiv
\frac{\dot{p}_N}{\dot{\rho}_N}\,, \eeq
where the dot differentiation here is with respect to whatever
definition of time. Notice that if the various components would have
an equation of state (EOS) of the form $p_N=w_N\,\rho_N$, with
constant EOS parameter $w_N$, then $c_{a,N}^2=w_N$. However, even in
this case the mixture has a variable effective EOS parameter, as we
will see in the next section.

On the other hand, if the perturbations are non-adiabatic, there is
an entropy contribution to the pressure perturbation\,\cite{MFB92}:
\begin{equation}
p_N\Gamma_N=\de p_N-c_{a,N}^2\de \rho_N\,,
\end{equation}
where $\Gamma_N\equiv(\delta p_N)_{\rm non-adiabatic}/p_N$ is the
intrinsic entropy perturbation of the $N$th component, representing
the displacement between hypersurfaces of uniform pressure and
uniform energy density \cite{Malik03}. {For covariantly conserved
components}, a gauge-invariant relationship between $\de p_N$ and
$\de \rho_N$ for a general non-adiabatic stress is given by
\cite{Malik03}-\cite{Kodama96}:
\begin{equation}\label{pnad} \de p_N=c_{s,N}^2\de \rho_N +
\,3Ha^2(\rho_N+p_N)(c_{s,N}^2-c_{a,N}^2)\frac{{\theta}_N}{k^2}\,,
\end{equation}
where $c_{s,N}^2$ can be regarded as a rest frame speed of sound. We
will refer to $c_{s,N}^2$ as the \emph{effective} speed of sound, in
the sense that we treat the cosmic fluid effectively as
hydrodynamical matter. Since (\ref{pnad}) is gauge invariant, the
perturbed quantities in this expression can be computed, in
particular, within the synchronous gauge. In this way, we can
consistently substitute (\ref{pnad}) in the equations
(\ref{PertEeq}), (\ref{EConsPert1}) and (\ref{EConsPert2}) to
eliminate the perturbation $\delta p_N$.  This allows us, finally,
to solve for the three basic sets of perturbations
\begin{equation}\label{indpvari2}
(\hh(t,k),\ \delta\rho_N(t,k),\ \theta_N(t,k))\,.
\end{equation}

\noindent \section{Perturbations for a composite DE fluid with a
variable effective equation of state } \label{sect:effEOS}

In this section, we apply the linear matter and dark energy density
perturbations to a general class of models in which the DE fluid is
a composite and covariantly self-conserved medium and matter is also
canonically conserved. From Eq.~(\ref{EnergyCons}), in the matter
dominated epoch ($p_M=0$), the matter component $\rM$ satisfies
\beq\label{MatCons} \rho_M^{\prime}(a) +\frac{3}{a} \rho_M(a)=0\,.
\eeq
Here we found convenient to trade the differentiation with respect
to the cosmic time ($^\circ$) for the differentiation with respect
to the scale factor. The latter is denoted by a prime (i.e.
$f'\equiv df/da$ for any $f$; hence $\dt{f}=a\,H\,f'$). The scale
factor is related to the cosmological redshift $z$ by $a(z)\equiv
1/(1+z)$, where we define $a(0)\equiv a_0=1$ at the present time. It
follows that the normalized matter density evolves as
\beq\label{rhoM} {\Omega_M}(a)={\Omega^0_M}a^{-3}\,, \eeq
where $\Omega_M^0$ is the corresponding current value. Since the
total DE is also globally conserved, from Eq.~(\ref{EnergyCons}) we
also obtain
 \beq \label{DECons}
\rho^{\prime}_D +\frac{3}{a} (1+\we)\rho_D=0 \,,
 \eeq
where $\we$ is the effective equation of state (EOS) parameter and
$\rD=\rho_1+\rho_2+...$ is the total density of the multicomponent
DE fluid. For a composite DE model, in which the DE is a mixture of
fluids with individual EOS $p_i=\omega_i\,\rho_i\ (i=1,2,...,n)$,
the effective EOS parameter is defined as
 \beq \label{mixture}
\we=\frac{p_D}{\rD}=\frac{\omega_1\,\rho_1+\omega_2\,\rho_2+...}{\rho_1+\rho_2+...}\,.
 \eeq

The Hubble expansion in terms of the normalized densities, in the
matter dominated period, follows from (\ref{friedeq}):
\beq\label{HLXCDM} H^2=\frac{8\pi G}{3}\rho_T=
H_0^2\,\left(\OMo\,a^{-3}+\OD(a)\right)\,, \eeq
where ${\Omega_D}$ is the normalized total DE density
${\Omega_D(a)}\equiv {\rD(a)}/{\rco}$.

The non-adiabatic perturbed pressure (\ref{pnad}) for the total DE
component can be written in terms of the effective EOS
as\,\footnote{In the particular case where the DE is a purely
running $\CC$, one has to consider the perturbations $\delta\rL$,
but since $\wL=-1$ it turns out that $\theta_D$ (here
$\theta_{\CC}$) plays no role and hence it is enough to consider the
relation $\de p_{\CC}=-\delta\rL$, see Ref.\,\cite{FabrisDens} for
details.}
 \beq\label{pnadDE}
{\de p_D}= \csd{\de
\rD}+3Ha^2(1+\we)\rD(\csd-\cad)\frac{\theta_D}{k^2}\,,
 \eeq
where we have omitted for simplicity the subindex `D' from the
adiabatic and effective speeds of sound of the DE (i.e. $\cad\equiv
c_{a,D}^2$, $\csd\equiv c_{s,D}^2$; this convention will be used
throughout the text). The units are taken to be the light speed $c =
1$ and $\hbar = 1$ such that the Planck scale is defined by $M_P =
G^{-1/2} = 1.22 \times 10^{19}$ GeV. In these units, usually $0\le
\csd\le 1$ for a general DE model. In this range, one can show that
for constant EOS parameter there is small suppression on the DE
fluctuation $\de\rD$ as $\csd$ increases\, \cite{Malik03}. Near-zero
(but not vanishing) sound speed today is possible in models like
k-essence, for example, in which the EOS parameter is positive until
the matter-dominated triggers a change to negative pressure; in this
kind of models, it is even possible to have $\csd > 1$, regime for
which the growth of the DE density perturbations is suppressed
\cite{Erickson01}.

In a non-perfect fluid, spatial inhomogeneities in $T_{\mu\nu}$
imply shear viscosity in the fluid. In this case, a possible
contribution to shear through a ``viscosity parameter" $c_{\rm
vis}^2$ should also be taken into account \cite{Hu98}. In principle,
$\csd$ is an arbitrary parameter. Nevertheless, the limit where
$(\csd, c_{\rm vis}^2)\to (1,0)$ corresponds exactly to a scalar
field component with canonical kinetic term \cite{Koivisto06}.

Using the total DE conservation law (\ref{DECons}), we can write the
total DE adiabatic sound speed (\ref{adspeed}) as
\begin{equation}\label{adspeedDE}
\cad\equiv \frac{{p}_D^{\prime}}{{\rho}_D^{\prime}}=
\we-\frac{a}{3}\frac{{w}_e^{\prime}}{(1+\we)} \,.
\end{equation}

The perturbed equations (\ref{EConsPert1}) and (\ref{EConsPert2})
for the (conserved) matter component (for which $p_M=\de p_M=0$) can
be written as differential equations in the scale factor:
 \beq \label{MPerta1}
{\de}_M^{\prime} &=&
-\frac{1}{aH}\left(\theta_M-\frac{\hh}{2}\right)\,,
\\ \label{MPertheta}
{\theta}_M^{\prime} &=& - \frac{2}{a}\theta_M \,,
 \eeq
where $\de_M\equiv{\de\rho_M}/{\rho_M}$ is the relative matter
fluctuation (density contrast). According to Eq.(\ref{MPertheta}),
the matter velocity gradient is decaying ($\theta_M\propto a^{-2}$).
Assuming the conventional initial condition $\theta_M^0\equiv
\theta_M(a=1)=0$, we have $\theta_M(a)=0\,(\forall a)$. So, the
perturbed matter set of coupled equations (\ref{MPerta1}) and
(\ref{MPertheta}) yields the simple relation
 \beq \label{hPert}
{\hh}=2aH{\de}^{\prime}_M\,. \eeq

Let us also define the relative fluctuation of the DE component,
$\de_D\equiv\de\rD/\rD$. Using the non-adiabatic perturbed pressure
(\ref{pnadDE}) and the DE conservation law (\ref{DECons}), we can
write the perturbed equations (\ref{EConsPert1}) and
(\ref{EConsPert2}) for the self-conserved DE fluid in the following
way
 \beq \label{DEPert}
{\de^{\,\prime}_D} &=& - \frac{(1+\we)}{aH}\left\{\left[
1+\frac{9a^2H^2(\csd-\cad)}{k^2} \right]
\theta_D-\frac{\hh}{2}\right\}
\nonumber\\
&& -\frac{3}{a}(\csd-\we)\de_D\,, \\
 \label{DEPertTheta}
{\theta}^{\prime}_D &=&-\frac{1}{a}\left(2-3\csd\right){\theta}_D+\frac{k^2}{a^3H
} \frac{\csd\de_D}{(1+\we)} \,,
 \eeq
where in the last equation we have used (\ref{adspeedDE}) to
eliminate $c_a^2$. Moreover, from (\ref{DEPertTheta}) one can see
that a negligible DE sound speed ($\csd\approx 0 $) causes the
velocity gradient to decay ($\theta_D\propto a^{-2}$), as in the
case of matter [Eq.(\ref{MPertheta})]. If we assume the conventional
initial conditions $\theta_M^0=\theta_D^0=0$, we have
$\theta_M=\theta_D=0\ (\forall a)$. In this case, the total DE fluid
is comoving with the matter as long as the Universe \text{and}
{perturbations} evolve, which is a very particular case. Actually,
for this case, the $k$ (scale) dependence disappears from the
equations. On the other hand, from Eq.(\ref{DEPertTheta}) one can
see that, if we neglect the DE perturbations, $\de\rD\approx 0$, for
a constant $\csd$ we obtain again $\theta_M=\theta_D=0$ and the
scale independence.

However, $\de\rD$ modifies the evolution of the metric fluctuations
according to the perturbed Einstein equation (\ref{PertEeq}); and,
in turn, this causes the corresponding evolution of the matter
perturbations through Eq.\,(\ref{hPert}). We can write down the
appropriate form of the perturbation equation as follows. First, we
define the ``instantaneous" normalized densities at a cosmic time
$t$, $\tilde{\Omega}_N={\rho_N}/{\rc}\,$, where $\rc=3H^2/8\pi\,G$
is the critical density at the same instant of time. (Notice that
the new densities equal the previously defined ones, i.e.
$\tilde{\Omega}_N=\Omega_N$, only at $t=t_0$.) Making use of the
definition of the relative DM fluctuation $\de_M$ and of the
non-adiabatic perturbed pressure (\ref{pnadDE}), we may cast
Eq.\,(\ref{PertEeq}) in the following way:
\beq \label{PertEeqDE}
{\hh}' + \frac{2}{a} \hh -  \frac{3H}{a} \tOM\de_M &=&
\frac{3H}{a}\tOD \Big[ (1+3\csd)\de_D \nonumber\\
&& +\, 9a^2
H(\csd-\cad)\frac{\theta_D}{k^2} \Big].
\eeq
Next we use Eq.(\ref{hPert}) to eliminate $\hh$ from
(\ref{PertEeqDE}). With the help of
\beq\label{Hdot1}
{H'(a)} &=& -\frac{4\,\pi\,G}{aH(a)}\,\left[\rM(a)+(1+\we(a))\,\rD(a)\right]
\\  \label{Hdot2}
&=& -\frac{3\,H(a)}{2a}\,\left[1+\frac{\we(a)\,r(a)}{1+r(a)}\right]
\eeq
and
\begin{equation}\label{tildeOMs}
\tilde{\Omega}_M (a)=\frac{1}{1+r(a)}\,,\ \ \ \ \ \ \ \ \ \
\tilde{\Omega}_D (a)=\frac{r(a)}{1+r(a)}\,,
\end{equation}
where $r(a)$ is the ``cosmic coincidence ratio'' (\ref{ra}) between
the DE and matter densities, we may finally rewrite
(\ref{PertEeqDE}) as follows
\begin{eqnarray}
\delta_M''(a)&+&\frac{3}{2}\left[1-\frac{{\we(a)}\,r(a)}{1+r(a)}
\right]\frac{\delta_M'(a)}{a}- \frac{3}{2}\frac{1}{1+r(a)}
\frac{\delta_M(a)}{a^2} \nonumber\\
&=&\frac{3}{2}\,
\frac{r(a)}{1+r(a)} \Big[(1+3\csd)\frac{\de_D(a)}{a^2} \nonumber\\
&& +\, 9\,\,H(a)\,(\csd-\cad)\frac{\theta_D(a)}{k^2} \Big]
\,.\label{ggrowth3}
\end{eqnarray}
{If we would neglect the DE fluctuations ($\de_D=0$, $\theta_D=0$),
the \textit{r.h.s.} of the previous equation would vanish. Under
these conditions, one would be left with a decoupled, second-order,
homogeneous differential equation that determines the matter
perturbations $\delta_M$. As could be expected, the resulting
equation coincides with Eq.\,(2.16) of Ref.\,\cite{GOPS}, where the
approximation of neglecting the DE perturbations was made right from
the start as an intermediate procedure to investigate the amount of
linear growth of the matter perturbations and put constraints on the
parameter space of the $\La$XCDM model. This procedure was called
``effective" in that reference, since all the information about the
DE is exclusively encoded in the non-trivial EOS function
$\we=\we(a)$. Let us write the homogeneous equation as follows,
\begin{equation}
\delta_M''(a)+ \frac{3}{2}(1-\we\,\tOD)\frac{\delta_M'(a)}{a}-
\frac{3}{2}(1-\tOD)\frac{\delta_M(a)}{a^2}=0\,, \label{ggrowth3ex}
\end{equation}
and let us assume a time interval not very large such that $\tOD$
and $\we$ remain approximately constant. Looking for a power-law
solution of (\ref{ggrowth3ex}) in the limit $\tOD\ll 1$, we find,
for the growing mode,
\begin{equation}\label{deltaOD}
\delta_M\sim a^{1-3(1-\we)\tOD/5}\sim
a\,\left[1-\frac{3(1-\we)}{5}\,\tOD\,\ln a\right]\,.
\end{equation}
Since $\we<0$ for any conceivable form of DE, this equation shows
very clearly that we should expect a growth suppression of matter
perturbations (i.e. $\delta_M\sim a^n$, with $n<1$) whenever a
(positive) DE density $\tOD$ is present within the horizon.
Physically speaking, we associate this effect to the existence of
negative pressure that produces cosmological repulsion of matter.}

However, being the DE density non-constant in general ($\delta_D\neq
0$), the DE perturbations themselves (and not only the value of the
background DE density) should act as a source for the matter
fluctuations. This effect is precisely encoded in the inhomogeneous
part of Eq.\,(\ref{ggrowth3}), i.e. in its \textit{r.h.s} which is,
in general, non-zero for $\delta_D,\theta_D\neq 0$. In order to
better appreciate this effect, let us consider another simplified
situation where the analytical treatment is still possible, namely,
let us assume an adiabatic regime ($\cs=\cad$) with roughly constant
EOS ($\we\simeq$ const.) at very large scales (for which $k$ in
Eq.\,(\ref{DEPertTheta}) is very small, and hence the $\theta_D$
component becomes negligible). Under these conditions,
Eq.\,(\ref{DEPert}) greatly simplifies as follows:
\begin{equation}\label{DEPert2}
\delta^{\prime}_D=\frac{(1+\we)}{a\,H}\,\frac{\hh}{2}=(1+\we)\,\delta_M^{\prime}\,,
\end{equation}
where in the second step we have used \,(\ref{hPert}). The rates of
change of the perturbations for matter and DE, therefore, become
proportional in this simplified setup. To be more precise, we see
from (\ref{DEPert2}) that for $\we\gtrsim -1$ (quintessence-like
behavior of the composite DE fluid) the matter fluctuations that are
growing with the expansion ($\delta_M^{\prime}>0$) trigger
fluctuations in the DE also growing with the expansion
($\delta^{\prime}_D>0$), whereas for $\we\lesssim -1$ (phantom-like
behavior of the DE) we meet exactly the opposite situation, i.e.,
increasing fluctuations in the matter density
($\delta_M^{\prime}>0$) lead to decreasing fluctuations in the DE
($\delta^{\prime}_D<0$). Note that upon trivial integration of
(\ref{DEPert2}) one finds $\delta_D=(1+\we)\,\delta_M+ C$, where $C$
is a constant determined by the initial conditions. For $C=0$ one
obtains a result that fits with the well-known adiabatic initial
condition relating the density contrasts of generic matter and DE
components\,\cite{cosmobooks},
\begin{equation}\label{adiabcond}
\frac{\delta_M}{1+w_M}=\frac{\delta_D}{1+w_D}\,,
\end{equation}
where, for non-relativistic matter, we have $w_M=0$, and $\wD$ is,
in this case, the effective EOS $\we$ of the composite DE fluid.
Since a positive DE density always leads to cosmological repulsion,
it follows from (\ref{DEPert2}) that one should expect some
inhibition (resp. enhancement) of the matter growth for the
quintessence-like (resp. phantom-like) case.

Although the previous example illustrates the impact of the DE
fluctuations on the matter growth for a simple situation, a more
complete treatment is required in the general case. In practice,
this means that we have to numerically solve the system
(\ref{hPert})-(\ref{PertEeqDE}) or, if desired, replace the last
equation with the second order inhomogeneous Eq.\,(\ref{ggrowth3})
whose \textit{r.h.s.} depends on the density contrast and the
velocity gradient for the DE, $\delta_D,\theta_D\neq 0$. Notice that
the presence of overdensity DE perturbations ($\delta_D>0$) does
\textit{not} necessarily imply the inhibition of the corresponding
matter perturbations since the coefficient $1+3\cs$ in front of
$\delta_D$ on the \textit{r.h.s.} of Eq.\,(\ref{ggrowth3}) is
positive for non-adiabatic DE perturbations. Only for $\csd=\cad$ we
meet the aforementioned possibility because $\cad\simeq \we$ is
usually negative, unless $\we$ is rapidly decreasing with the
expansion, see Eq.\,(\ref{adspeedDE}). In this sense, the discussion
above, based on Eq.\,(\ref{DEPert2}), is only valid at very large
scales, specifically for $k$-modes whose length scale $\ell\sim
k^{-1}$ is outside the sound horizon (cf. section
\ref{sect:genericfeatures})\,\footnote{In section~\ref{DEfluct}, we
will see that this particular situation does accommodate very well
the realistic dynamics of the cosmic perturbations for matter and DE
during the early stages of the evolution.}. However, at smaller
scales, specially at scales inside the sound horizon, and for a
general non-adiabatic regime, we need to solve the aforesaid
complete system of equations for the basic set of perturbation
variables for the metric, matter and DE:
\begin{equation}\label{setvariables}
(\hh(a,k),\ \delta_M(a,k),\ \delta_D(a,k),\ \theta_D(a,k))\,.
\end{equation}
In this way, we have extended the effective treatment of the DE
perturbations presented in Ref.\,\cite{GOPS}, and we are now ready
to better assess the scope of its applicability. In section
\ref{sect:perturbLXCDM}, we will apply this general formalism to the
$\CC$XCDM model.

\section{Some generic features of the DE
perturbations}\label{sect:genericfeatures}

In the present section, we summarize some characteristic features of
the DE perturbations. Many properties which are, in principle,
common to any model with a self-conserved DE, will be later
exemplified in section \ref{DEfluct} within the non-trivial context
of the so-called $\CC$XCDM model\,\cite{LXCDM12,GOPS}.

\subsection{Divergence at the CC boundary}

In general, the EOS of the DE will be a dynamical quantity,
$\we=\we(a)$. In many models, the EOS may change from
quintessence-like ($-1<\we<-1/3$) to phantom ($\we<-1$) behavior or
vice versa, acquiring therefore the value $\we=-1$ (also referred to
as the `CC boundary') at some instant of time. This is problematic
since, as we shall see next, the equations for the perturbations
diverge at that point.

The divergence at the CC boundary is common to any DE model and has
been thoroughly studied in the literature (see e.g.
\cite{Vikman04,Caldwell05}). The problem can be readily seen by
direct inspection of Eqs. (\ref{DEPert}) and (\ref{DEPertTheta}).
Note that, even though $\cad$ diverges at the crossing (cf.
(\ref{adspeedDE})), the combination $(1+\we)\,\cad\ $ remains finite
and, therefore, Eq.(\ref{DEPert}) is well-behaved. Thus, the problem
lies exclusively in the $(1+\we)$ factor in the denominator of
(\ref{DEPertTheta}). One might think that the divergence can be
absorbed through a redefinition of the variables, but this is not
the case.

Getting around this difficulty is not always possible. It is
well-known that there is no way for a single scalar field model to
cross the CC boundary\,\cite{Vikman04}. The simplest way to avoid
this problem is to assume two fields $(Q,P)$, e.g. one that works as
quintessence $w_Q>-1$ and dominates the DE density until the
CC-crossing point, and beyond it the other field retakes the
evolution with a phantom behavior $w_P<-1$, or the other way around;
see \cite{Caldwell05,2field} for a detailed discussion and specific
parameterizations of $\we$. As we will see, in the $\La$XCDM model
the additional restrictions needed to avoid this divergence will
further constrain the physical region of the parameter space.

\subsection{Unbounded growth for adiabatic DE perturbations}

Another well-known problem is the unbounded growth of the DE
perturbations for a negative squared speed of sound $\csd$.
As already mentioned in the previous section, in the adiabatic case we have
$\cad\simeq\we$, which is in general negative as long as the EOS
parameter is not varying too fast. As a result, the adiabatic DE
perturbations may lead to explosive growth unless extra degrees of
freedom are assumed (see e.g. \cite{Koivisto06} for discussion).

In order to better see the origin of the problem, let us rewrite
Eqs.(\ref{DEPert}) and (\ref{DEPertTheta}) in terms of conformal
time $\eta$, which is easily done by making use of $\dot{f}=
a^2\,H\,f'$ (for any $f$):
 \beq
\dot{\de}_D &=& -\, a(1+\we)\left\{\left[
1+\frac{9\mathcal{H}^2(\csd-\cad)}{k^2} \right]
\theta_D-\frac{\hh}{2}\right\}
\nonumber\\
&& -\,3 \mathcal{H}(\csd-\we)\de_D\,,\\
\dot{\theta}_D &=&-\, \mathcal{H}\left(2-3\csd\right){\theta}_D+\frac{k^2}{a}
\frac{\csd\de_D}{(1+\we)} \,.
 \eeq
As in section \ref{sect:perturbations}, we have defined
$\mathcal{H}=\dot{a}/a\equiv Ha$. If we use the two equations above
and Eq.\,(\ref{PertEeqDE}) to get a second order differential
equation for $\de_D$, we arrive at
\begin{equation}\label{2ndorder}
\ddot{\de}_D=-k^2\csd\de_D+{\cal O}(\hat{h},
\delta_M,\theta_D,\de_D)\,,
\end{equation}
where the second term on the \textit{r.h.s.} represents other terms
linear in these variables. Assuming that the various perturbations
are initially more or less of the same order, we see that the first
term on the \textit{r.h.s.} of (\ref{2ndorder}) will dominate
provided
\begin{equation}\label{shcond}
\left|k\int c_{s} \mathrm{d}\eta\right|\gg1\,.
\end{equation}
Notice that, for constant sound velocity, this condition simply
tells us that the wavelength of the modes satisfies $\ell\sim
k^{-1}\ll\lambda_{s}$, where $\lambda_{s}=c_{s}\eta$ is the ``sound
horizon''. Eq. (\ref{shcond}) is a generalization of this condition
for arbitrary sound speed, in which case the sound horizon is given
by
\begin{equation}\label{sh0}
\lambda_{s}=\int^{\eta}_0 c_{s} \mathrm{d}\eta=\int^a_0
\frac{c_{s}\,\mathrm{d}\tilde{a}}{\tilde{a}^2\,H(\tilde{a})}\,,
\end{equation}
and constitutes a characteristic scale for the DE perturbations. As
we will see next, the DE is expected to be smooth for scales well
below it \cite{Hu98, Takada06}.

For scales well inside the sound horizon, (\ref{2ndorder}) becomes
the equation of a simple harmonic oscillator, whose solution is (in
what follows, we assume constant $\csd$ for simplicity):
\begin{equation}
\de_D=C_1 e^{i c_{s}k\eta}+C_2 e^{-i c_{s} k\eta}\,,
\end{equation}
where $C_1$ and $C_2$ are constants. We see that, for $\csd<0$ (i.e.
imaginary $c_s$) and neglecting the decaying mode, the DE
perturbations grow exponentially. Obviously, this situation is
unacceptable for structure formation\,\footnote{Let us note that
this kind of problem need not occur when the DE is a pure running
$\CC$\,, for then the perturbation $\delta\rL$ is no longer an
independent dynamical variable and is, in fact, determined by an
algebraic function of the other perturbation variables, see
\cite{FabrisDens} for details.}. On the other hand, if $\csd>0$ the
DE density contrast oscillates. Since $\de_M$ grows typically as the
scale factor $a$, this ensures that the ratio $\de_D/\de_M\sim
\de_D/a\to 0$ with the expansion. In other words, this tells us that
the DE is going to be a smooth component (as it is usually assumed)
as long as we are well inside the sound horizon. This feature is
treated in more depth in the following section.

\subsection{Smoothness of DE below the sound horizon}

As a matter of fact, Eq.(\ref{2ndorder}) is an oversimplification.
In addition of having a term proportional to $\de_D$, we also have
one depending on its first derivative. So we can write that equation
more precisely as follows:
\begin{equation}
\ddot{\de}_D=D_1\,\de_D+D_2\,\dot{\de}_D+{\cal
O}(\hat{h},\delta_M,\theta_D)\,,
\end{equation}
which gives us not just a simple, but a damped harmonic oscillator.
The coefficients $D_1$ and $D_2$ are, in general, functions of the
conformal time $\eta$. So the DE density contrast does not only
oscillate, but its amplitude decreases with time. Indeed, it was
shown in \cite{Hu98} that the quantity
\begin{equation}
\de_D^{\rm (rest)}=\de_D+3\mathcal{H}a(1+\we)\frac{\theta_D}{k^2}\,,
\end{equation}
which corresponds to the density contrast in the DE rest frame,
oscillates with an amplitude $A$ that decreases according to
\begin{equation}
A\propto c_{s}^{-1/2} a^{(-1+3\we)/2}\,.
\end{equation}

The damped oscillations of the DE density contrast are clearly seen
in the $\CC$XCDM model, as we will show in section \ref{DEfluct}.

Finally, we may ask ourselves whether the scales relevant to the LSS
surveys\,\cite{Cole05} are well inside the sound horizon or not.
Note that, in a matter dominated Universe with negligible CC term
and constant $c_s$, we have $H^2=H_0^2\,\OMo\,a^{-3}$ and the sound
horizon (\ref{sh0}) takes the simple form
\begin{equation}\label{sh}
\lambda_{s}=\frac{2\,c_{s}}{H_0\,(\OMo)^{1/2}}\sqrt{a}\,.
\end{equation}
Thus, in general, we expect that the size of the sound horizon at
present ($a_0=1$) should be roughly of the order of the Hubble
length $H_0^{-1}$. On the other hand, the observational data
concerning the linear regime of the matter power spectrum lie in the
range $0.01h {\rm Mpc}^{-1}<k<0.2h {\rm Mpc}^{-1}$\,\cite{Cole05},
which corresponds to length scales $\ell\sim k^{-1}$ comprised in
the interval ${(600\,H_0)^{-1}}<\ell<{(30\,H_0)^{-1}}$, hence well
below the sound horizon (at least for $\csd$ not too close to 0).
Therefore, according to the previous discussions, we expect the DE
density to be smooth at those scales, and indeed it will be so for
the $\CC$XCDM model. Nevertheless, as we will see in section
\ref{DEfluct}, the larger the scale $\ell$ or the smaller the speed
of sound $c_s$, the more important the DE perturbations are, because
then (\ref{shcond}) is not such a good approximation.

\noindent \section{The $\CC$XCDM model as a candidate to solve the
cosmic coincidence problem} \label{sect:LXCDM}

The $\Lambda$XCDM model\,\cite{LXCDM12} provides an interesting way
of explaining the so-called ``cosmological coincidence problem" (cf.
section \ref{sect:cosmiccoincicence}). The idea is related to the
possibility of having a dynamical component $X$, called the
``cosmon"\,\footnote{Originally, the cosmon appeared in
Ref.\,\cite{PSW} as a scalar field linked to the mechanism of
dynamical adjustment of the CC. In the present context, the entity
$X$ is also differentiated from the CC, but if taken together they
form a composite and interactive DE medium.}, which interacts with a
running cosmological constant $\Lambda$. If the matter components
are canonically conserved, the composite DE ``fluid'' made out of
$X$ and the running $\CC$ will be a self-conserved medium too. The
dynamics of the $\Lambda$XCDM universe is such that its composite DE
may enforce the existence of a stopping point after many Hubble
times of cosmological expansion. As a result, this modified
FLRW-like universe can remain for a long while in a situation where
the coincidence ratio (\ref{ra}) does not change substantially from
the time when the DE became significant until the remote time in the
future when the stopping point is attained. Subsequently, the
Universe reverses its motion till the Big Crunch.

The total DE density and pressure for the $\Lambda$XCDM universe are
obtained from the sum of the respective CC and $X$ components:
\beq\label{tDEX} \rho_D = \rho_{\Lambda}+\rho_X\,, \qquad p_D =
p_{\La}+p_X\,.
 \eeq
The evolving CC density $\rho_{\Lambda}(t)=\CC(t)/8\pi\,G$ of the
model is motivated by the quantum field theory formulation in curved
space-time by which the CC is a solution of a renormalization group
equation. Following \cite{JHEPCC1,SSIRGAC06,ShS08,fossil,RGTypeIa},
the CC density emerges in general as a quadratic function of the
expansion rate:
 \beq \label{runlamb}
\rho_\La(H)\,=\,\rho_\Lambda^0 + \frac{3\,\nu}{8\pi}\,M_P^2\,
\left(H^2-H^2_0\right)\,,
 \eeq
where $\rho_\Lambda^0=\rL(H=H_0)$ is the present value. The
dimensionless parameter $\nu$ is given by
 \beq \label{nu}
\nu\equiv \frac{\sigma}{12\,\pi}\,\frac{M^2}{M_P^2}\,,
 \eeq
where $M$ is an effective mass parameter representing the average
mass of the heavy particles of the Grand Unified Theory (GUT) near
the Planck scale, after taking into account their multiplicities.
Depending on whether they are bosons, or fermions, $\sigma = +1$, or
$\sigma = -1$, respectively. For example, for $M=M_P$ one has
$|\nu|=\nu_0$, where
 \beq \label{nu0}
\nu_0\equiv\frac{1}{12\pi}\simeq 2.6\times 10^{-2}\,.
 \eeq
On physical grounds, we expect that this value of $|\nu|$ should be
the upper bound for this parameter. In the next section, we will see
if we can pinpoint a region of parameter space compatible with this
expectation.

The energy density associated to the cosmon component $X$ is
obtained from the total DE conservation law (\ref{DECons}) and the
composite form (\ref{tDEX}),
 \beq \label{LXCDMCons}
\rho_{X}^{\prime}(a)+\rho_{\La}^{\prime}(a) = -\frac{3}{a}\, (1+w_X)
\rho_X(a)\,,
 \eeq
where $w_X$ is the effective EOS parameter of $X$,
 \beq \label{EOSX}
p_X\equiv w_X\rho_{X}\,.
 \eeq
In principle, $w_X$ could be a function of the scale factor.
However, a simpler assumption that allows us to perform a completely
analytic treatment, is to consider that the $X$ component behaves as
a barotropic fluid with a constant EOS parameter in one of the
following two ranges: $\wX \gtrsim -1$ (quintessence-like cosmon) or
$\wX \lesssim -1$ (phantom-like cosmon). On the other hand, the EOS
parameter for the running $\La$ component still remains as the
cosmological constant one, $w_{\La}=-1$, i.e.
 \beq \label{EOSL}
p_{\La}\equiv -\rho_{\La}\,.
 \eeq
From these assumptions, it is easy to find the following relation
between the effective EOS parameter of the total DE fluid
(\ref{mixture})
and the EOS parameter of the cosmon $w_X$:
 \beq \label{dewx}
\left(1+\we(a)\right) \rho_D(a)=\left(1+w_X\right) \rX(a)\,.
 \eeq
The normalized density of the cosmon component,
$\OX(a)=\rX(a)/\rho_c^0$, can be obtained from the previous
relations {after solving the differential equation
(\ref{LXCDMCons}). In this equation, we have
$\rL^{\prime}(a)=(3\nu/8\pi)\,M_P^2\,dH^2/da$ from (\ref{runlamb}),
and the derivative $dH^2/da=2\,H(a)\,H'(a)$ can be explicitly
computed from (\ref{Hdot1}) upon using (\ref{dewx}) and
(\ref{rhoM}). One finally obtains the differential equation
\begin{equation}\label{OXdiff}
\OX^{\prime}(a)+\frac{3}{a}\,\left(1+\wX-\epsilon\right)\,\OX(a)=3\,\nu\,\OMo\,a^{-4}\,.
\end{equation}
With the boundary condition that the current value of $\rX$ is
$\rX^0$, the solution of (\ref{OXdiff}) can be written in the
following way}:\,\footnote{The fact that the evolution of the cosmon
$X$ is completely determined by the dynamics of the running $\rL$
(\ref{runlamb}), together with the hypothesis of total DE
conservation (\ref{LXCDMCons}), implies that $X$ cannot be generally
assimilated to a scalar field, which has its own dynamics. In fact,
as we have already mentioned, $X$ is to be viewed in general as an
effective entity within the context of the effective action of QFT
in curved space-time. }
 \beq \label{rhoX}
\OX(a)=\OXo\,\left[ \left({1+b}\right)\,
a^{-3(1+w_X-\epsilon)}-b\,a^{-3}\right]\,,
 \eeq
where in the above equations we have used the notations
 \beq
b &\equiv& -\ \frac{\nu\,\OMo}{(\wX-\epsilon)\,\OXo}\,,\label{b}
\\ \nonumber
\\ \label{epsilon}
\epsilon &\equiv&\nu(1+w_X)\,.
 \eeq
As we will discuss in more detail below, the parameter $\epsilon$
must remain small ($|\epsilon|< 0.1$) in order to be compatible with
primordial nucleosynthesis. For $\nu=0$ the CC density
(\ref{runlamb}) becomes constant. In this case, the two parameters
(\ref{b}) and (\ref{epsilon}) vanish and Eq.\,(\ref{rhoX}) boils
down to the simplest possible form, which is characteristic of a
self-conserved monocomponent system,
 \beq \label{rhoXsimplest}
\OX(a)=\OXo a^{-3(1+w_X)}\,.
 \eeq
It is only in this particular situation where the cosmon $X$ could
be a self-conserved scalar field with its own dynamics. But in
general this is not so because in QFT in curved space-time we have
good reasons to expect a running
$\rL$\,\cite{JHEPCC1,SSIRGAC06,ShS08}, and hence $\nu\neq 0$.
Therefore, if the total DE is to be conserved, the dynamics of $X$
is not free anymore and becomes determined as in (\ref{rhoX}).

The normalized total DE density ${\Omega_D}=\rD/\rco$ is given by
 \beq \label{normDE}
{\Omega_D} (a)&=& \left(\frac{1-\OLo}{1-\nu}
-\frac{w_X\Omega^0_{M}}{w_X-\epsilon}\right)a^{-3(1+w_X-\epsilon)}
\nonumber\\
&& \,+\, \frac{{\Omega_{\Lambda}^0}-\nu} {1-\nu}
+\frac{\epsilon\,\Omega^0_{M}\,a^{-3}}{w_X-\epsilon} \,,
\eeq
where the various current normalized densities satisfy the relation
$\OMo+\ODo=\OMo+\OLo+\OXo=1$, which may be called the ``$\CC$XCDM
cosmic sum rule''. Using (\ref{dewx}), the effective EOS of the DE
in the $\CC$XCDM model can now be obtained explicitly,
\begin{equation}\label{weLCDM}
\we(a)=-1+(1+\wX)\,\frac{\OX(a)}{\OD(a)}\,,
\end{equation}
with $\OX(a)$ and $\OD(a)$ given by (\ref{rhoX}) and (\ref{normDE}),
respectively.

The total DE density (\ref{normDE}) varies in such way that the
ratio (\ref{ra}) can remain under control, which is the clue for
solving the coincidence problem in a dynamical way\,\cite{LXCDM12}.
Indeed, the explicit computation of such ratio yields
 \beq \label{ratioLXCDM}
r(a)&=& \left[\frac{1-\OLo}{\OMo\,(1-\nu)}\,
-\frac{w_X}{w_X-\epsilon}\right]\,a^{-3(w_X-\epsilon)} \nonumber\\
&& \,+ \, \frac{(\OLo-\nu)\,a^{3}} {(1-\nu)\,\Omega^0_{M}}
+\frac{\epsilon\,}{w_X-\epsilon}
\,,
 \eeq
and it can be bounded due to the existence of a maximum (triggered
by the $\sim a^{-3(w_X-\epsilon)}$ term in the previous formula).
Moreover, $r(a)$ given above stays relatively constant (typically
not varying more than one order of magnitude) for a large fraction
of the history of the Universe and for a significant region of the
parameter space\,\cite{LXCDM12}. In contrast, in the standard
concordance $\La$CDM model, the CC density remains constant,
$\rL=\rLo$, and the coincidence ratio grows unstoppably with the
cubic power of the scale factor,
$r(a)=\Omega_{\La}^0\,a^3/\Omega_M^0 $. In this scenario, it is
difficult to explain why the constant $\rL=\rLo$ is of the same
order of magnitude as the matter density right now: $\rho_{M}^0$.
Let us point out that the standard model ratio is just that
particular case of (\ref{ratioLXCDM}) for which $\nu=0$ (no running
CC) and $\OXo=0$ (no cosmon).

Before closing this section, we would like to make a remark and some
discussion concerning the quadratic evolution law (\ref{runlamb})
for the cosmological term. This equation was originally motivated
within the framework of the renormalization group (RG) of QFT in
curved space-time\,\cite{cosm,JHEPCC1,Babic,Gruni,RGTypeIa} (see
also\,\cite{SSIRGAC06} for a short review). We point out a criticism
against this approach that recently appeared in the
literature\,\cite{AGS}. While the nature of this criticism was amply
rejected in \cite{ShS08} (see below for a summary), it is fair to
say that the question of whether a rigorous  RG approach in
cosmology is feasible is still an open question and remains a part
of the CC problem itself. Although it is not our main aim here to
focus on this fundamental issue, let us briefly sketch the situation
along the lines of Ref.\,\cite{ShS08}, to which we also refer the
reader for a summary of the rich literature proposing different RG
formulations of cosmology both in QFT in curved space-time and in
Quantum Gravity.

The RG method in cosmology treats the vacuum energy density as a
running parameter and aims at finding a fundamental differential
relation (renormalization group equation) of the form
\begin{equation}\label{RGCC}
\frac{d\rL}{d\ln\mu}=\beta_{\La}(P, \mu)\,,
\end{equation}
which is supposed to describe the leading quantum contributions to
it, where $\beta_{\La}$ is a function of the parameters $P$ of the
effective action (EA) and $\mu$ is a dimensional scale. The
appearance of this arbitrary mass scale is characteristic of the
renormalization procedure in QFT owing to the intrinsic breaking of
scale invariance by quantum effects. The quantity $\rL$ in
(\ref{RGCC}) is a ($\mu$-dependent) renormalized part of the
complete QFT structure of the vacuum energy. Depending on the
renormalization scheme, the scale $\mu$ can have a more or less
transparent physical meaning, but the physics should be completely
independent of it. Such (overall) $\mu$-independence of the
observable quantities is actually the main message of the RG; but,
remarkably enough, the $\mu$-dependence of the individual parts is
also the clue of the RG technique to uncover the leading quantum
effects.

Essential for the RG method in cosmology is to understand that, in
order for the vacuum energy to acquire dynamical properties, we need
a nontrivial external metric background. The dynamical properties of
this curved background (e.g. the expanding FLRW space-time,
characterized by the expansion rate $H$) are expected to induce a
functional dependence $\rL=\rL(H)$. The latter should follow from
parameterizing the quantum effects with the help of the scale $\mu$
and then using some appropriate correspondence of $\mu$ with a
physical quantity, typically with $H$ in the cosmological context,
although there are other possibilities\,\cite{cosm,JHEPCC1,Babic}.
In this way, one expects to estimate the subset of quantum effects
reflecting the dynamical properties of the non-trivial background.
Although $\mu$ cancels in the full EA, the RG method enables one to
separate the relevant class of quantum effects responsible for the
running. The procedure is similar to the RG in a scattering process
in QCD; the parametrization of the quantum effects in terms of $\mu$
is the crucial strategy to finally link them with the energy of the
process through the correspondence $\mu\to q$ (where $q$ is a
typical momentum of the scattering process) at high energy. One can
also proceed in the same way in QED and electroweak theory (although
here one can adopt more physical subtraction schemes, if desired).
The RG technique can actually be extended to the whole Particle
Physics domain. In cosmology, however, the situation is more
complicated, partly because (as remarked above) the physical scale
behind the quantum effects is not obvious. Still, one expects that
it should be related with the expanding metric background and hence
the expansion rate $H$ can be regarded as a reasonable possibility.
On this basis, the heuristic arguments exhibited in
\cite{JHEPCC1,ShS08,Babic,Gruni,fossil,RGTypeIa} combined with the
general covariance of the EA suggest that the solution of the RG
equation (\ref{RGCC}) should lead to the kind of quadratic law
(\ref{runlamb}) that we have used.

According to \cite{ShS08}, the point of view of Ref.\,\cite{AGS} is
incorrect on two main accounts: first, because they try to disprove
the running through the overall cancelation of the arbitrary scale
$\mu$ in the EA; and second, because they neglect the essential role
played by the non-trivial metric background. As emphasized in
\cite{ShS08}, the cancelation of $\mu$ in the EA cannot be argued as
a valid criticism because this fact is a built-in feature of the RG
and it was never questioned. If this would be a real criticism, it
would also apply to QED, QCD or any other renormalizable QFT, and
nevertheless this is no obstacle for using the RG method in these
theories as an extremely useful strategy to extract the dependence
of the quantum effects on the physical energy scale of the
processes, in particular the so-called running coupling constants
$g_s=g_s(q)$ and $e=e(q)$ of the strong and electromagnetic
interactions. Moreover, in the absence of a non-trivial metric
background, there is no physical running of the vacuum energy, even
though there is still $\mu$-dependence of the various parts of the
EA and in particular of the CC, see e.g. \,\cite{Brown}. Therefore,
at the end of the day such criticism seems to go against the essence
of the RG method and its recognized ability to encapsulate the
leading quantum effects on the physical observables.

In cosmology, the principles of the RG should be the same, but there
are two main stumbling blocks that prevent from straightforwardly
extending the method in practice\,\cite{ShS08}, to wit: i) the
aforesaid lack of an obvious/unique correspondence of $\mu$ with a
cosmological scale defining the physical running, and also (no less
important) ii) the huge technical problems related with the
application of the RG within a physical (momentum-dependent)
renormalization scheme in a curved background. These difficulties
are unavoidable here because we are dealing with QFT in the infrared
regime and moreover the metric expansions cannot be performed on a
flat background; indeed, there cannot be a flat background in the
presence of a cosmological term! While these two problems remain
unresolved in a completely satisfactory manner, it is legitimate to
use the phenomenological approach and the educated guess (e.g. the
general covariance of the EA) to hint at the running law. This is
the guiding principle followed in the aforementioned references and
that led to Eq.\,(\ref{runlamb}).

Finally, let us emphasize that irrespective of whether such law can
be substantiated within the strict framework of the RG, the present
study remains perfectly useful simply treating (\ref{runlamb}) as an
acceptable type of a phenomenological variation law and keeping also
in mind that adding the cosmon may contribute to the resolution of
the pressing cosmic coincidence problem.

\section{Dark energy perturbations in the $\CC$XCDM model}\label{sect:perturbLXCDM}

In this section, we further elaborate on the conditions to bound the
ratio $r=r(a)$ and discuss the constraints on the parameter space,
in particular the impact of the DE perturbations on these
constraints. In section \ref{sect:genericfeatures}, we discussed
analytically some generic features about DE perturbations. In
principle, those results should apply to any model in which the DE
is self-conserved. The $\CC$XCDM model, given its peculiarities (a
composite DE which results in a complicated evolution of the
effective EOS), constitutes a non-trivial example of that kind of
models. In this sense, it is interesting to use the $\CC$XCDM  to
put our general predictions to the test. At the same time, this will
allow us to impose new constraints on the parameter space of the
model, improving its predictivity.

The parameter space of the $\CC$XCDM model was already tightly
constrained in \cite{GOPS}. In that work, the matter density
fluctuations were analyzed under the assumption that the DE
perturbations could be neglected. As a first approximation this is
reasonable since, as we have discussed in section
\ref{sect:genericfeatures}, the DE is expected to be smooth at
scales well below the sound horizon. Thus, we will take the results
of \cite{GOPS} as our starting point and will check numerically the
goodness of that approximation. Finally, we will further constrain
the parameter space using the full approach presented in the present
work. Let us summarize the constraints that were imposed in
\cite{GOPS}:

\begin{enumerate}
\item {Nucleosynthesis bounds}:
As already commented, the ratio (\ref{ratioLXCDM}) between DE and
matter densities should remain relatively small at the
nucleosynthesis time, in order not to spoil the Big Bang model
predictions on light-element abundances. Requiring that ratio to be
less than  $10\%$, it roughly translates into the condition
$|\epsilon|<0.1$, where the parameter $\epsilon$ was defined in
Eq.\,(\ref{epsilon}).

\item {Solution of the coincidence problem}: In Ref.\, \cite{LXCDM12}, where
the $\CC$XCDM model was originally introduced as a possible solution
to the coincidence problem, it was shown that there is a large
sub-volume of the total $\CC$XCDM parameter space for which the
ratio $r(a)$ remains bounded and near the current value $r_0$ (say,
$|r(a)|\lesssim10\,r_0$, where $r_0\sim 7/3$) during a large
fraction of the history of the Universe. Thus, the fact that the
matter and DE densities are comparable right now may no longer be
seen as a coincidence. Such solution of the coincidence problem is
related to the existence of a future stopping (and subsequent
reversal) of the Universe expansion within the relevant region of
the parameter space.

\item {Current value of the EOS parameter}: Recent studies (see eg. \cite{WMAP3})
suggest that the value of the DE effective EOS should not be very
far from $-1$ at present. Although these results usually rely on the
assumption of a constant EOS parameter (and thus are not directly
applicable to the $\CC$XCDM model), we adopted a conservative point
of view and stuck to them by enforcing the condition
$|1+\we(a=1)|\leq0.3$ on the EOS function (\ref{weLCDM}).

\item {Consistency with LSS data}: As said before, in \cite{GOPS}
we studied the growth of matter density fluctuations under the
assumption that the DE was smooth on the scales relevant to the
linear part of the matter power spectrum. From the fact that the
standard $\CC$CDM model provides a good fit to the observational
data, we took it as a reference and imposed that the amount of
growth (specifically, the matter power spectrum), of our model did
not deviate by more than $10\%$ from the $\CC$CDM value (``F-test''
condition). This condition can be also be justified from the
observed galaxy fluctuation power spectrum, see \cite{GOPS} for more
details.
\end{enumerate}

The upshot of that analysis was that there is still a big sub-volume
of the three dimensional $\CC$XCDM parameter space
$(\nu,\,\wX,\,\OLo)$ satisfying simultaneously the above
conditions\,\footnote{In Ref.\cite{GOPS}, we took a prior for the
normalized matter density at present, specifically $\OMo=0.3$. This
means that $\ODo=0.7$ for a spatially flat Universe. For better
comparison with those results, we keep this prior also in the
present work.}. The projections of that volume onto the three
perpendicular planes $(\nu,\OLo)$, $(\nu,\,\wX)$ and $(\OLo,\wX)$
are displayed in Figs.\,\ref{fig1} and \ref{fig2} (shaded regions).
These regions where already determined in Ref.\,\cite{GOPS}. In the
next section, we will discuss how the final set of allowed points
becomes further reduced when we take into account the analysis of
the DE perturbations.

\subsection{Divergent behavior at the CC boundary} \label{parspace}

As discussed in section \ref{sect:genericfeatures}, if the effective
EOS of the model crosses the CC boundary ($\we=-1$) at some point in
the past, the perturbation equations will present a real divergence.
Obviously, this circumstance makes the numerical analysis unfeasible
at the points of parameter space affected by the singularity.

In the absence of an apparent mechanism to get around this
singularity, we are forced to restrict our parameter space to the
subregion where the solution of the perturbation equations
(\ref{hPert})-(\ref{PertEeqDE}) is regular, namely by removing those
points of the parameter space that present such a crossing in the
past, because these points can not belong to a well defined history
of the Universe. In the absence of a more detailed definition of the
cosmon entity $X$, this new constraint is unavoidable. This should
not be considered as a drawback of the model, for even in the case
when one uses a collection of elementary scalar fields to represent
the DE, one generally meets the same kind of divergent behavior as
soon as the CC boundary is crossed, unless some special conditions
are arranged. In other words, even if the components of the DE are
as simple as, say, elementary scalar fields with smooth behavior and
well-defined dynamical properties (including an appropriately chosen
potential), there is no a priori guarantee that the CC boundary can
be crossed safely\,\cite{Vikman04}. {It is possible to concoct
ingenious recipes, see e.g. \,\cite{Caldwell05}, such that the
perturbation equations become regular at the CC boundary, but the
procedure is artificial in that one must introduce new fields (one
quintessence-like and another phantom-like) satisfying special
properties such that their respective EOS behaviors match up
continuously at the CC-crossing. Apart from the fact that fields
with negative kinetic terms are not very welcome in QFT, one cannot
just replace the original fields with the new ones without at the
same time changing the original DE model!} As we will see below, in
the $\CC$XCDM case the absence of CC-crossing projects out a region
of the parameter space which is significantly more reduced, and
therefore the predictive power of the model becomes substantially
enhanced.

\begin{figure}
\begin{center}
\resizebox{\columnwidth}{!}{\includegraphics{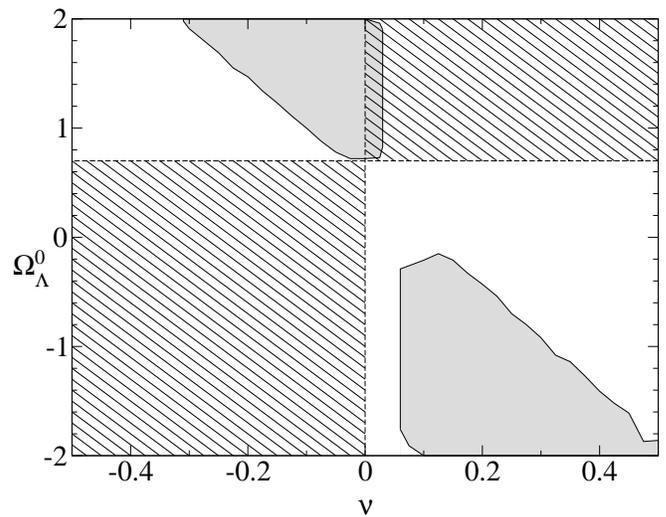}}
\caption{Projection of the 3D physical volume of the $\CC$XCDM model onto the $\nu-\OLo$
plane. Shaded area: points that satisfy all the constraints in
\cite{GOPS}, see also section \ref{sect:perturbLXCDM} of the present
work. Striped area: points that are not affected by the divergence
at the CC boundary discussed in Sect.~\ref{parspace}. The final
allowed region is the one both shaded and stripped. As a result of
considering the DE perturbations, the possible values of the
parameters become strongly restricted, which implies a substantial
improvement in the predictive power of the model.} \label{fig1}
\end{center}
\end{figure}

\begin{figure}
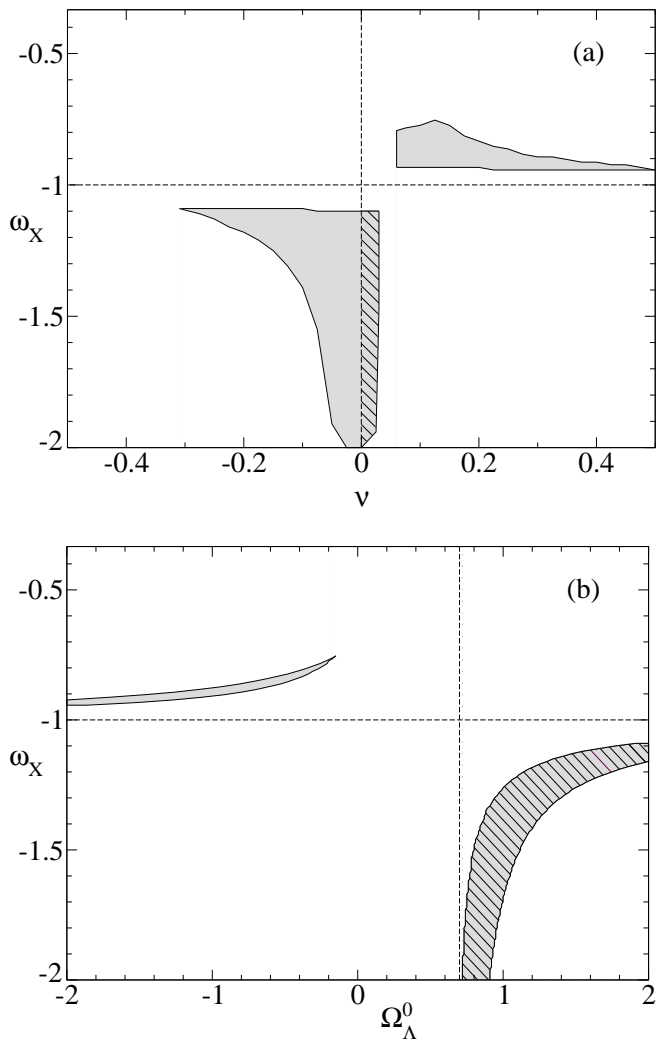

\resizebox{\columnwidth}{!}{\includegraphics{Figures/fig2a.eps}}\\[3ex]
\resizebox{\columnwidth}{!}{\includegraphics{Figures/fig2b.eps}}
\caption{Projection of the 3D physical volume of the $\CC$XCDM model onto the $\nu-\wX$
and $\OLo-\wX$ planes. Shaded: points that satisfy all the
constraints in \cite{GOPS}. Striped: points that, in addition, are
not affected by the divergence at the CC boundary discussed in Sect.
\ref{parspace}. The final allowed region is the one both shaded and
stripped.} \label{fig2}
\end{figure}

In section 6 of the first reference in\,\cite{LXCDM12}, it was shown
that the necessary and sufficient condition for having a CC boundary
crossing in the past within the $\CC$XCDM model is that the
parameter $b$ given in (\ref{b}) is positive. As can be readily
seen, this will happen whenever $\nu$ and $\Omega_X^0$ have the same
sign (where we use the fact that $\wX<0$ and $|\epsilon|\ll|\wX|$ in
the relevant region of parameter space). From the cosmic sum rule of
the $\CC$XCDM model, we have $\OXo=1-\OMo-\OLo$; thus, using our
prior $\OMo=0.3$, the set of allowed points (for which $\OXo$ has
\textit{different} sign from $\nu$) are those comprised in the
striped areas in Figs.\,\ref{fig1} and \ref{fig2}. This leaves us
with a very small allowed region in each plane, which is just the
corresponding intersection of the shaded area and the striped one.

At the end of the day, it turns out that most of the points in the
shaded area in Figs.\,\ref{fig1} and \ref{fig2} (viz. those allowed
by the conditions stated in the previous section and the analysis of
\cite{GOPS}) are ruled out by the new constraint emerging from the
DE perturbations analysis, and hence we end up with a rather
definite prediction for the values of the $\CC$XCDM parameters. In
particular, we find from these figures that only small positive
values of $\nu$ are allowed, at most of order $\nu\sim 10^{-2}$. Let
us emphasize that this is in very good agreement with the
theoretical expectations mentioned in section \ref{sect:LXCDM}.
Recall that, from the point of view of the physical interpretation
of $\nu$ in Eq.\,(\ref{nu}), we expected $\nu$ in the ballpark of
$\nu_0\sim 10^{-2}$ at most -- see  Eq.\,(\ref{nu0}) -- since the
masses of the particles contributing to the running of the CC should
naturally lie below the Planck scale\,\footnote{Let us clarify that
the tighter bounds on $\nu$ determined in Ref.\,\cite{FabrisDens}
are possible only because, in the latter work, the DE is not
conserved and there is no cosmon. As we have shown in \cite{GOPS}, a
running cosmological constant model without a self-conserved DE
cannot solve the coincidence problem in a natural way because the
required values of $\nu$ are too large and, hence, incompatible with
the physical interpretation of this parameter. }. Let us mention
that the interesting bounds on $\nu$ obtained in
Ref.\,\cite{HorvatGSL} on the basis of the so-called generalized
Second Law of gravitational thermodynamics would suggest that only
the effective mass near the Planck mass is allowed. However, let us
point out that such study has been performed without including the
non-trivial effect from the cosmon.


A very important consequence of the dark energy perturbative
constraint is that the effective EOS of the DE can be
quintessence-like only, i.e. $-1<\we<-1/3$. To prove this statement,
let us start from Eq.\,(\ref{dewx}). For the current values of the
parameters, this equation can be rewritten as
\begin{equation}\label{aux}
\big(1+\we^0\big)\,\ODo=(1+\wX)\,\OXo\,,
\end{equation}
where $\we^0\equiv\we(a=1)$ is the value of the effective EOS
parameter at the present time. {Looking at Figs.\,\ref{fig2}a and
\ref{fig2}b, we realize the following two relevant features: first,
the cosmon component is necessarily phantom-like ($\wX<-1$) in the
allowed region by the DE perturbations; and, second, its energy
density at present is negative; namely, $\OXo=0.7-\OLo<0\ $ because
from Fig.\,\ref{fig2}b we have $\ \OLo>0.7$}. Therefore, since the
\textit{r.h.s.} of (\ref{aux}) is constrained to be positive and
$\ODo = 0.7>0$, we are enforced to have $\we^0>-1$. However, the
fulfilment of this condition at present implies its accomplishment
in the past, i.e. $\we(a)>-1 \ (\forall a\leqslant 1)$, otherwise
there would have been a crossing of the CC boundary at some earlier
time, which is excluded by the analysis of the DE perturbations. The
upshot is that the EOS of the DE in the $\CC$XCDM model can only
appear effectively as quintessence (\textit{q.e.d}). In reality, it
only mimics quintessence, of course, as its ultimate nature is
\textit{not}; such DE medium is a mixture made out of running vacuum
energy and a compensating entity that insures full energy
conservation of the compound system.

{Dark energy components $X$ with negative energy density are
peculiar in cosmology since, in contrast to standard DE components,
they satisfy the strong energy condition (like ordinary matter), and
as a result the gravitational behavior of $X$ is attractive rather
than repulsive. Due to this double resemblance with matter and
phantom DE (although with the distinctive feature $\rX<0$), such
components can be called ``phantom matter'' \,\cite{LXCDM12}}. Being
$X$ in general an effective entity, such ``phantom matter'' behavior
is actually non-fundamental.

\subsection{Adiabatic speed of sound in the $\CC$XCDM}

In the equations (\ref{hPert})-(\ref{PertEeqDE}), we assumed the
most general case in which the perturbations could be non-adiabatic.
Moreover, we have shown that the adiabatic case usually leads to an
unphysical exponential growth of the perturbations as a result of
$\cad$ in (\ref{adspeedDE}) being negative. Next we will check that,
indeed, the most common situation in the $\CC$XCDM model is to have
$\cad<0$. Notwithstanding, adiabatic perturbations are not
completely forbidden in the present framework, as there is a small
region of the parameter space for which $\cad$ could be positive.

From Eqs.\,(\ref{adspeedDE}) and (\ref{dewx}), and making use of the
DE conservation law (\ref{DECons}), we find that the adiabatic speed
of sound for the $\CC$XCDM model can be cast as follows,
\begin{equation}
\cad (a)= - 1 - \frac{a}{3}\
\frac{{{\Omega}_X^{\prime}}(a)}{{\Omega}_X(a)}\,.
\end{equation}
With the help of (\ref{rhoX}), we can rewrite the last expression as
\begin{eqnarray}
\label{adspeedX} \cad (a)&=&\frac{\OXo}{\OX(a)}(1+b)(\wX-\epsilon)\
a^{-3(1+\wX-\epsilon)}\,.
\end{eqnarray}
We want to find out the condition for this expression to be
positive. As $(\wX-\epsilon)<0$ (remember that $\wX<0$ and
$|\epsilon|<0.1$, due to nucleosynthesis constraints), that
condition simply reads
\begin{equation}
\frac{\OXo}{\OX(a)}(1+b)<0\,.
\end{equation}
The cosmon energy density $\OX(a)$ cannot vanish because, in such
case, the perturbation equations would diverge. Indeed, $\OX(a)=0$
corresponds to a CC-boundary crossing at some value of the scale
factor in the past, cf. Eq.\,(\ref{weLCDM}). Thus, being $\OX(a)$ a
continuous function, it must have the same sign in the past as that
of its present value, i.e. $\OXo/\OX(a)>0\ (\forall a\leqslant 1))$.
In short, the final condition that ensures that $\cad>0$ is
\begin{equation}
(1+b)<0\label{cond}\,.
\end{equation}
If this condition would be satisfied, then $\cad>0$ would hold for
the entire past history of the Universe and, under these
circumstances, the adiabatic equations may be used. It turns out
that the relation (\ref{cond}) can be satisfied in the $\CC$XCDM
model, although only in a narrow range of the parameter space. In
fact, from the definition of the parameter $b$ in (\ref{b}), the
expectation that $|\wX|={\cal O}(1)$, and neglecting $\epsilon$, we
see that (\ref{cond}) is approximately equivalent to
\begin{equation}
\OXo\gtrsim \frac{\nu\,\OMo}{\wX}={\cal O}(-\nu)\,.
\end{equation}
Given the fact that $\nu$ was found to be positive and small  (cf.
Fig.\,\ref{fig1}-\ref{fig2}) and, at the same time, $\OXo<0$ (see
previous section), the above condition does not leave much freedom
within the allowed parameter space, roughly $-\nu\lesssim\OXo<0$.
This narrow strip is, however, not necessarily negligible; e.g. if
we take $\nu$ of order of $\nu_0\sim 10^{-2}$ (cf.
Eq.\,(\ref{nu0})), this possibility is still permitted in the
parameter space, see Figs.\,\ref{fig1}-\ref{fig2}. In such case, the
present cosmon density could still be of the order or larger (in
absolute value) than, say, the current neutrino contribution to the
energy density of the Universe ($\Omega^0_{\nu}\sim 10^{-3}$). No
matter how tiny is (in absolute value) a negative cosmon
contribution to the energy density, it suffices to take care of the
cosmic coincidence problem along the lines that we have explained.
Therefore, the adiabatic contribution is perfectly tenable, but the
numerical analysis of the subsequent sections remains essentially
the same (as we have checked) independently of whether the sound
speed of the DE medium is adiabatic or not. For this reason, in what
follows we will assume the more general situation of non-adiabatic
perturbations, with the understanding that adiabatic ones can do a
similar job in the corresponding region of the parameter space.


\section{The matter power spectrum}\label{numerical}

In this section, we compare the matter power spectrum predicted by
the $\CC$XCDM model with the observed galaxy power spectrum measured
by the 2dFGRS survey \cite{Cole05}. The $\CC$XCDM matter power
spectrum is found by evolving the perturbation equations
(\ref{hPert})-(\ref{PertEeqDE}) from $a=a_i$ to the present
($a_0=1$), where $a_i\ll 1$ is the scale factor at some early time,
but well after recombination. In these equations, we must of course
use the expansion rate (\ref{HLXCDM}) with the full DE density
(\ref{normDE}).

In order to set the initial conditions at $a=a_i$, we use the
prediction from the standard $\CC$CDM model. Indeed, the standard
$\CC$CDM model provides a good analytical fit to the 2dFGRS observed
galaxy power spectrum. Taking this fit as our starting point, we
compute analytically the values of the $\CC$CDM perturbations at an
arbitrary scale factor. Since the DE does not play an important role
until very recently, we may assume that the initial matter and
metric perturbations at $a=a_i$ for the $\CC$XCDM model are the same
as for the $\CC$CDM model.


\subsection{Initial matter and metric perturbations}\label{icsection}

As previously commented, the perturbed equations are evolved from
some $a_i\ll 1$ to $a_0=1$. The value of $a_i$ is unimportant,
provided that it lies well after recombination (to insure that all
the processes encoded in the transfer function have already taken
place). For definiteness, we take $a_i=1/500$, i.e. cosmological
redshift $z\sim 500$. Next we specify the initial conditions at
$a=a_i$.

For the matter and metric perturbations, the initial conditions in
the standard cosmological model can be computed
analytically\,\cite{cosmobooks,MFB92}. The matter perturbed
equations in the standard $\CC$CDM model are (\ref{hPert}) and
(\ref{ggrowth3}) taking $\delta_D(a)=\theta_D(a)=0$. In these
conditions, the \textit{r.h.s.} of the differential equation
(\ref{ggrowth3}) vanishes and we are left with an homogeneous
equation. In the standard model, this equation is just
Eq.\,(\ref{ggrowth3ex}) with $\we=-1$, i.e.
\begin{equation}
\delta_M''(a)+ \frac{3}{2}\,(2-\tOM(a))\frac{\delta_M'(a)}{a}-
\frac{3}{2}\,\tOM(a)\frac{\delta_M(a)}{a^2}=0\,. \label{ggrowthSM}
\end{equation}
Since matter is conserved, we have
$\tilde{\Omega}_M(a)=(H_0/H(a))^2\,\OMo/a^3$. Moreover, from
Eq.\,(\ref{Hdot2}) (again with $\we=-1$) one can also see that
$\tOM(a)=-(2a/3)\,H'(a)/H(a)$. Hence, the homogeneous equation
(\ref{ggrowthSM}) can be conveniently cast as follows:
\begin{equation}\label{homog}
\delta_M''(a)+\left(\frac{3}{a}+\frac{H'(a)}{H(a)}\right)\,\delta_M'(a)-\frac{3}{2}\,\Omega_M^0\,
\frac{H_0^2}{H^2(a)}\,\frac{\delta_M(a)}{a^5}=0\,.
\end{equation}
This equation is now in a standard form and can be solved
analytically\,\cite{Dodelson}. Let us first introduce the variable
$D(a)={\de_M(a)}/{\delta_{\rm ref}}$ (the so-called growth factor),
where $\delta_{\rm ref}$ is the matter density contrast at some
initial scale factor. In the initial matter era ($a=a_i\ll 1$)
wherein the cosmological term is negligible, we have $H^2(a)/H^2_0
=\OMo\,a^{-3}$ in very good approximation; then, by imposing the
boundary conditions $D(a)\propto a$ and $D'(a)=1$, the growing
solution of (\ref{homog}) is simply $D(a)=a$, as can be easily
checked. In the general case, one can find the solution for the
growing mode which reduces to the previous one deep into the matter
dominated era. The final result reads
\begin{equation}\label{Da}
D(a)=\frac{\de_M(a)}{\delta_{\rm ref}}=\frac{5\Omega_M^0}{2}
\frac{H(a)}{H_0}\int^a_0\frac{d\tilde{a}}{(\tilde{a}\,H(\tilde{a})/H_0)^3}\,.
\end{equation}
In particular, for $H(a)/H_0 =(\OMo)^{1/2}\,a^{-3/2}$ it just boils
down to the solution $D(a)=a$ corresponding to the early matter
dominated epoch, as expected.

Furthermore, Eq.~(\ref{hPert}) leads to
\begin{equation}\label{initial}
\hh(a)=2aH(a)\frac{D'(a)}{D(a)}\de_M(a)\,,
\end{equation}
and for $D(a)=a$ it renders the initial condition
\begin{equation}
\hh(a_i)=2H(a_i)\de_M(a_i)\label{metricci}
\end{equation}
for the metric fluctuation. Later on, when the DE (i.e. $\rLo>0$ in
the $\CC$CDM) starts to play a role, the matter (and metric)
fluctuations become suppressed. The suppression is given by the
value of the growth factor $D(a)$, which is no longer proportional
to the scale factor\,\footnote{Notice that if
$\tilde{\Omega}_{\Lambda}$ is small and essentially constant, then
the growth factor $D(a)$ takes the approximate form $D(a)\sim a^n$,
with $n=1-6\,\tOL/5<1$, as it follows from (\ref{deltaOD}) for
$\we=-1$, or from (\ref{ggrowthSM}). This demonstrates, if only
roughly, the suppression behavior in an explicit analytic way. In
the general $\CC$CDM case, however, the solution for the growing
mode is given by (\ref{Da}), in which $H$ is the full expansion rate
of the standard model.}. From (\ref{Da}) it is clear that
$\delta_M(a)/D(a)$ is a constant, which can be written as
$\delta_M(a_i)/a_i$ at early times (when $D(a_i)=a_i$) and as
$\delta_M(a_0)/D(a_0)$ at the present time. Therefore,
\begin{eqnarray}\label{inic1}
\de_M(a_i)&=&\frac{a_i}{D_0}\de_M(a_0)\,,\\
\label{inic2} \hh(a_i)&=&2H_{\CC}(a_i)\frac{a_i}{D_0}\de_M(a_0)\,,
\end{eqnarray}
where $D_0\equiv D(a_0)$ {and the subindex in $H_{\CC}(a_i)$ has
been added to emphasize that the initial value of the Hubble
parameter is to be computed within the $\CC$CDM model. Note that,
instead of setting the value of $\hh(a_i)$, we could have chosen to
put initial conditions on the derivative of the density contrast,
$\delta_M'$. In that case, as it is evident from (\ref{inic1}), we
would have that $\delta_M'(a_i)=\delta_M(a_0)/D_0$. Then the initial
value of the metric fluctuation is constrained by Eq. (\ref{hPert})
to be $\hh(a_i)=2H_{\CC\rm X}(a_i) a_i\de_M(a_0)/D_0$, where now
$H_{\CC \rm X}(a_i)$ is the $\CC$XCDM value of the Hubble function.
Note that this value of $\hh(a_i)$ is not exactly the same as that
in (\ref{inic2}), since $H_{\CC}(a_i)$ and $H_{\CC\rm X}(a_i)$ are
not identical. However, being the difference rather small (as we
have checked numerically), the behavior of the perturbations does
not depend significantly on that choice}.

{The equations (\ref{inic1}) and (\ref{inic2})} give us the initial
conditions at $a_i=1/500$ for the matter and metric perturbations in
terms of the density contrast today $\de_M(a_0)$. We associate the
latter with the 2dFGRS observed galaxy power spectrum fitted in the
$\Lambda$CDM model, as detailed below.

The matter power spectrum of the $\CC$CDM model can be approximated
as \cite{FabrisDens}:
\begin{equation}\label{Powers}
P_{\CC}(k) \,\equiv\ |\de_M(k)|^2 \,=\,
A\,\,k\,\,T^2(k)\,\frac{g^2(\Om_T^0)}{g^2(\Om^0_{M})}\,,
\end{equation}
where $\Om_T^0=\OMo+\OLo$. It assumes a scale-invariant
(Harrison-Zeldovich) primordial spectrum, as generically predicted
by inflation. This primordial spectrum is modified when taking into
account the physical properties of different constituents of the
Universe, in particular the interactions between them. All these
effects are encoded into the scale-dependent transfer function
$T(k)$, which describes the evolution of the perturbations through
the epochs of horizon crossing and radiation/matter transition. The
growth at late times which, in the $\CC$CDM model, is independent of
the wavenumber, is described by the growth function $g(\Omega)$.
Finally, $A$ is a normalization factor.

The transfer function can be accurately computed by solving the
coupled system formed by the Einstein and the Boltzmann equations.
Although a variety of numerical fits have been proposed in the
literature, here we use the so-called BBKS transfer function
\cite{BBKS}:
 \beq \label{jtf}
T(k)&=&\frac{\ln (1+2.34 q)}{2.34 q}\Big[1+3.89 q + (16.1 q)^2\nonumber\\
&& + \, (5.46 q)^3+(6.71 q)^4\Big]^{-1/4}\,, \eeq
where
\begin{equation}
q=q(k)\equiv\frac{k}{(h \Ga)\,{\rm Mpc}^{-1}}\,
\end{equation}
and $\Ga$ is the Sugiyama's shape parameter
\cite{Sugiyama95,cosmobooks}
\begin{equation}
\Ga\,\equiv\,\Om_{M}^0\,h\,
e^{-\,\Om_B^0\,-\sqrt{{h}/{0.5}}\,\left({\Om_B^0}/{\Om_{M}^0}\right)
} \,.
\end{equation}
On the other hand, for the growth function we assume the following
approximation \cite{Carroll92}:
\begin{equation}
\label{CPT} g(\Om)=\frac{5\Om}{2}\, \left[\Om^{4/7} - \Om_{\La}^0 +
\Big(1+\frac{\Om}{2}\Big)
\Big(1+\frac{\Om_\La^0}{70}\Big)\right]^{-1}\,,
\end{equation}
which reflects the suppression in the growth of perturbations caused
by a positive cosmological constant. The normalization coefficient
$A$ is related to the CMB anisotropies through \cite{FabrisDens}
\begin{equation}
A\,=\, (2l_H)^4\, \frac{6\pi^2}{5}\,\frac{Q_{rms-PS}^2}{T_0^2} \,,
\label{AA}
\end{equation}
where $Q_{rms-PS}$ is the quadrupole amplitude of the CMB anisotropy
(see below for more detailed explanations), $l_H\equiv
H_0^{-1}\simeq 3000h^{-1}$Mpc is the Hubble radius and $T_0\simeq
2.725$K, the present CMB temperature. Therefore, the value of the
normalization factor $A$ could in principle be inferred from
measurements of the CMB. However, we have obtained it by fitting the
power spectrum (\ref{Powers}) to the 2dFGRS observed galaxy power
spectrum \cite{Cole05}, as discussed below.




We assume $h=0.7$ for the reduced Hubble parameter and a spatially
flat Universe with $\Omega_M^0=0.3$ (hence $\Omega_{\Lambda}^0=0.7$
for the flat $\CC$CDM model) in order to be consistent with our
assumption in previous analyses \cite{LXCDM12,GOPS}. The fit to the
2dFGRS observed galaxy power spectrum $P_{\rm 2dF}(k)$ is obtained
assuming the matter budget composed of a baryonic part
$\Omega_B^0=0.04$ and a dark matter contribution
$\Omega_{DM}^0=0.26$. In order to calculate the best fit, we use the
formula (\ref{Powers}) to minimize the $\chi$-square distribution
\begin{equation}
\chi^2\equiv \frac{1}{n_{dof}}\sum_k \frac{[P_{\rm
2dF}(k)-P_{\CC}(k)]^2}{\sigma^2(k)}
\end{equation}
in terms of the normalization $A$. There are 39 values of $k$ in the
2dFGRS data, so the number of degrees of freedom is $n_{dof}=38$.
We find, as best
fit, the value
\begin{equation}\label{AAfit}
A=8.99\times 10^5  h^{-4}{\rm Mpc}^4\,
\end{equation}
with $\chi^2=0.43$. From (\ref{AA}) we see that this value of $A$
implies $Q_{rms-PS}\simeq 20.85\;\mu K$. Let us now clarify that
$Q_{rms-PS}$ is not the \emph{observed} quadrupole CMB anisotropy
(usually denoted $Q_{rms}$), but rather the value derived from a fit
to the entire CMB power spectrum (PS).  For a power-law spectrum
with $n=1$ (i.e. for a scale-invariant PS), the COBE team obtained
$Q_{rms-PS}=18\pm1.6\;\mu K$ \cite{Bennett}, and moreover they found
that the observed $Q_{rms}$ is smaller than the fitted $Q_{rms-PS}$.
Whether this is a chance result of cosmic variance or reflects the
physical cosmology is not known\,\cite{Bennett}. Our fitted value
for $Q_{rms-PS}$ falls within the $2\sigma$ range of the
corresponding COBE value (although when the quadrupole itself is not
used in the fit, the COBE uncertainties become larger
\cite{Bennett}). As several authors have noted \cite{Sugiyama95,
Stompor}, such a normalization may be inadequate for models with a
cosmological constant, given the fact that the CMB spectra of
$\Lambda$-dominated models is quite different from a simple
power-law, specially at large scales (low multipoles). For instance,
in \cite{Stompor} it is proposed an alternative normalization for
the $\CC$CDM model, which for $h=0.8$ and $\OLo=0.7$ yields
$Q_{rms-PS}=22.04\;\mu K$, with an error of the order of $11\%$,
which is in agreement with our result. In general one can find a
number of different values for $Q_{rms-PS}$ in the literature
depending on the kind of analysis performed or the data set used,
and this is why we preferred to compute the normalization directly
from a fit to the matter power spectrum. Finally, let us emphasize
that our value for $Q_{rms-PS}$ lies within the $95\%$ confidence
interval for the  observed  quadrupole anisotropy ($Q_{rms}$) by
both COBE \cite{Bennett} and WMAP \cite{Hinshaw}.

Therefore, we will assume (\ref{inic1}) and (\ref{inic2}) as the
initial conditions for the matter and metric perturbations,
identifying $\de_M(a_0)$ with $\de_M(k)$ from the formula
(\ref{Powers}) using the fitted coefficient (\ref{AAfit}) .


\subsection{The ${\CC\rm X}$CDM matter power spectrum}

The procedure discussed above helps us to set initial conditions for
the matter and metric perturbations in the ${\CC\rm X}$CDM model.
However, since the $\CC$CDM model does not include DE perturbations,
we should set independent initial conditions on $\delta_D(a_i)$ and
$\theta_D(a_i)$. As already discussed, the scales relevant to the
matter power spectrum remain always well below the sound horizon
(\ref{sh}) and we expect negligible DE perturbations at any time.
Thus, the most natural choice for the initial values of the DE
perturbations is:
\begin{eqnarray}\label{DEic}
\delta_D(a_i)=0\,,\,\,\,\,\,\,\,\,\,\,\theta_D(a_i)=0.
\end{eqnarray}

{Indeed, this is not the only reasonable choice. For instance, we
could have also assumed the adiabatic initial condition
(\ref{adiabcond}) for the DE density contrast, i.e.
$\delta_D(a_i)=(1+\we(a_i))\delta_M(a_i)$, with $\delta_M(a_i)$
given by Eq.(\ref{inic1}). Again, it has been checked that the
evolution of the perturbations does not depend significantly on the
particular initial condition used.}

{Assuming the initial conditions (\ref{inic1}),(\ref{inic2}) and
(\ref{DEic}) for the matter, metric and DE perturbations at $a_i$,
respectively}, we can solve the perturbed equations
(\ref{hPert})-(\ref{PertEeqDE}). Equivalently, we can solve
(\ref{DEPert}),(\ref{DEPertTheta}) to obtain $(\delta_D,\theta_D)$
and then (\ref{ggrowth3}) to get the matter density fluctuations
today $\de_M(k,a=1)$ for any dark energy model. In particular, we
can (as a consistency check) solve the perturbation equations for
the $\CC$CDM model (in that case $\delta_D(a)=\theta_D(a)=0$, so the
only equations needed are (\ref{hPert}) and (\ref{PertEeqDE})). In
doing so, we recover exactly the spectrum $P_{\CC}$ defined in
(\ref{Powers}).

Now we proceed to compute the spectrum of the $\CC$XCDM, $P_{\CC\rm
X}(k)$. In order to better compare the shape of the different
spectra and the goodness of their fit to the 2dFGRS observed galaxy
power spectrum, we will normalize them at the {smallest length scale
considered $\ell\sim k^{-1}$, i.e. at $k=0.2$}, taking the $\CC$CDM
spectrum (\ref{Powers}) as reference. To this purpose, we introduce
a normalization factor $A_{\CC\rm X}$ in the matter power spectrum:

\begin{eqnarray}\label{pklx}
P_{\CC\rm X}(k)\equiv A_{\CC\rm X} |\delta_M(k,a=1)|^2\,.
\end{eqnarray}

Notice that the normalization factor $A_{\CC\rm X}$ gives us the
difference in the matter power spectrum of the model with respect to
that of the $\CC$CDM at the specific scale $k=0.2$. {Let us clarify
that the reason for choosing this scale for the normalization is
that, as discussed in section \ref{sect:genericfeatures}, the
smaller the length scale (i.e. the higher the value of $k$) the less
important the DE perturbations are. Therefore, at $k=0.2$, the
matter power spectrum of the model should not depend significantly
on whether we consider the effect of the DE perturbations; in
particular, it should be independent of the speed of sound $c_s$.
For larger scales, however, the DE perturbations can be more
significant and, as we shall see below, they may introduce some
differences in the shape of the power spectrum, which are
nevertheless small in the linear regime.}

\begin{figure}[t]
\resizebox{\columnwidth}{!}{\includegraphics{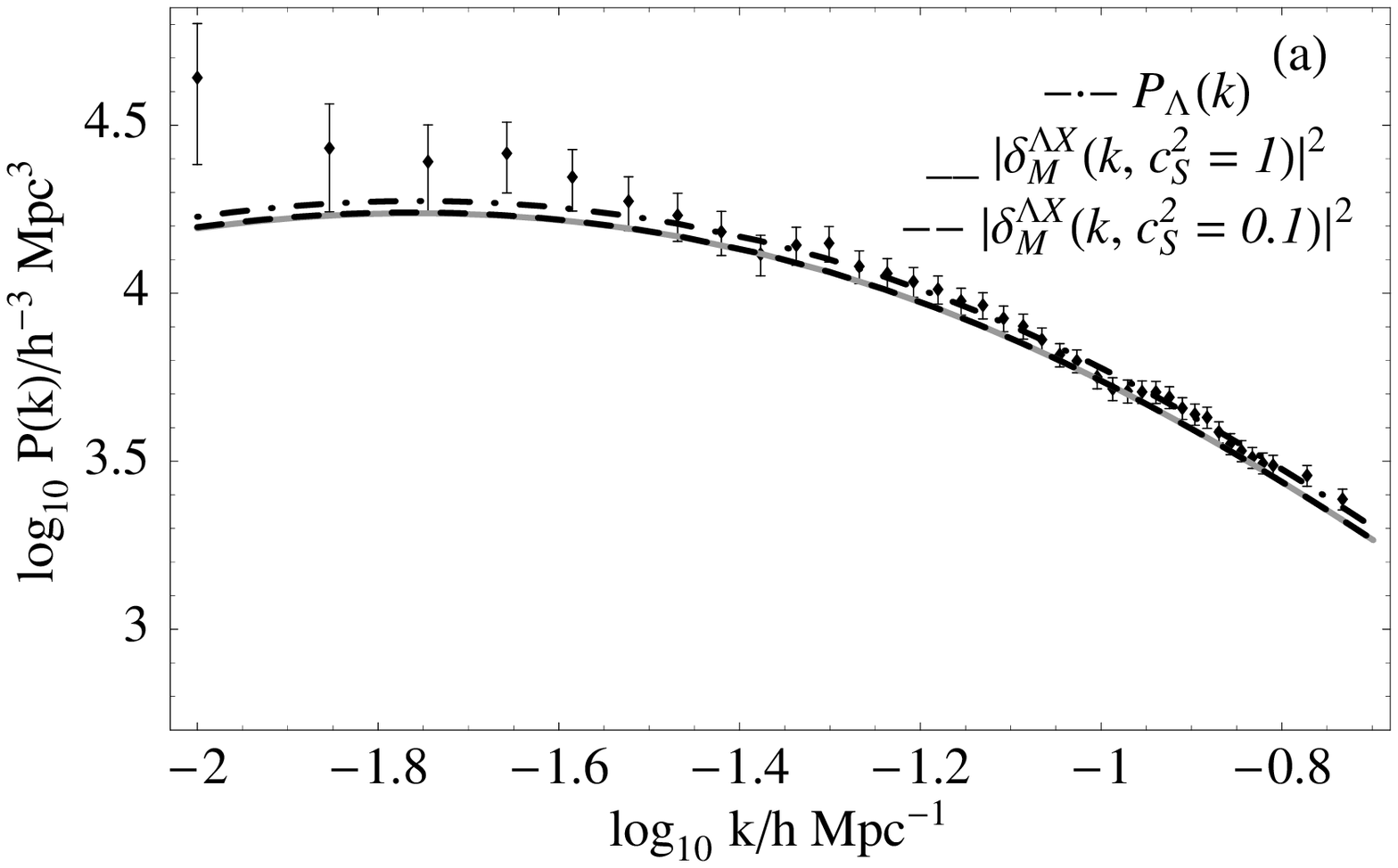}}\\[3ex]
\resizebox{\columnwidth}{!}{\includegraphics{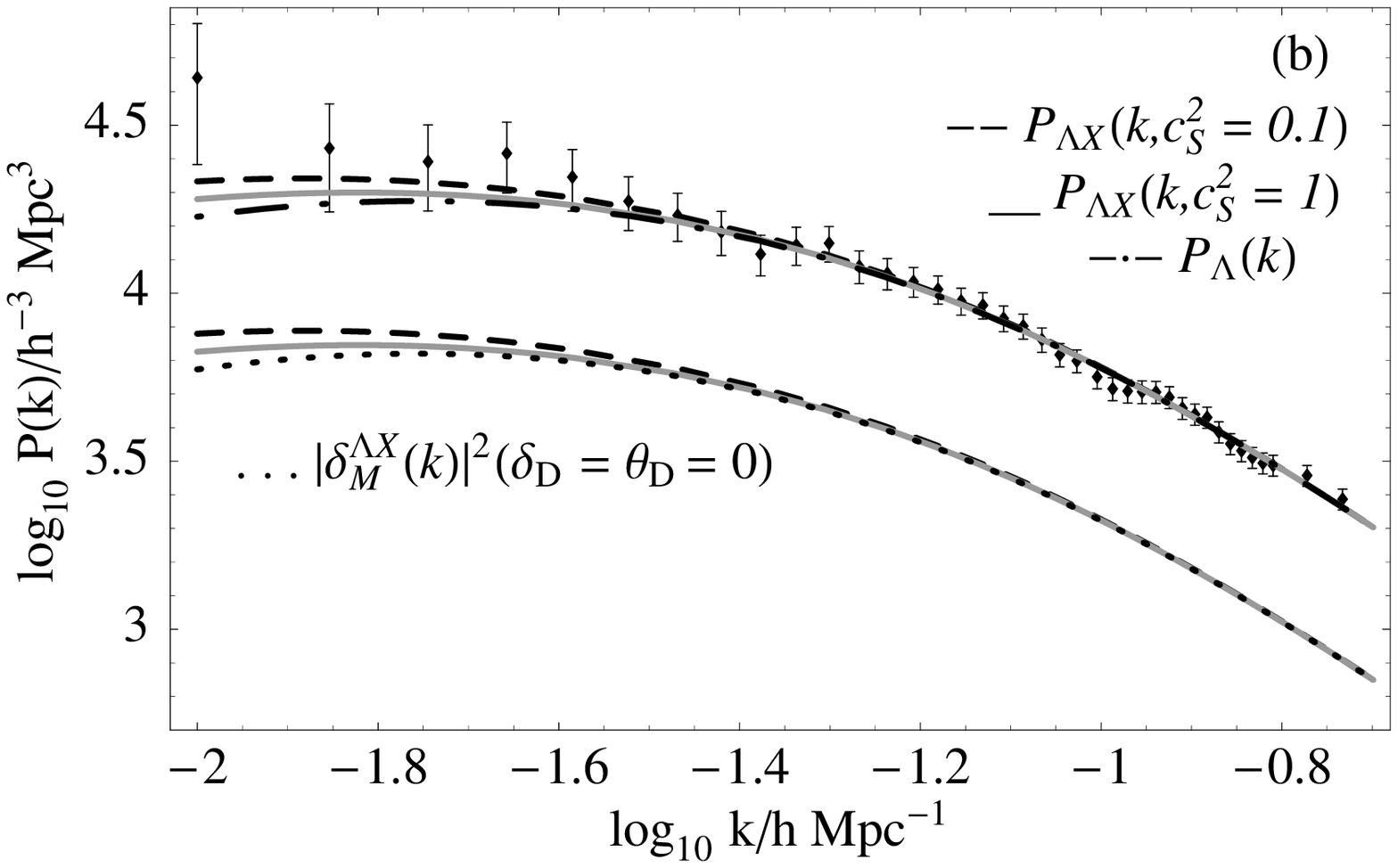}}
\caption{The $\CC$CDM power spectrum $P_{\Lambda}(k)$ (dot-dashed
line) versus the normalized and unnormalized spectrum predicted by
the $\CC$XCDM model, for DE sound speeds $c_{s}^2=0.1$ (dashed line)
and $c_{s}^2=1$ (solid/gray line): (a) for a set of parameters
allowed by the analysis of Ref.\cite{GOPS} (cf. shaded and striped
region in our Figs.~\ref{fig1}-\ref{fig2}), $\Omega_{\Lambda}^0 =
0.8$, $\nu = \nu_0\equiv 2.6\times 10^{-2}$ and $w_X=-1.6$. The
corresponding curves $P_{\Lambda}(k)$ and $P_{\Lambda \rm{X}}(k)$
coincide in this case; (b) for a set of parameters not allowed by
the F-test\,\cite{GOPS} (points in the stripped, but non-shaded,
region in our Fig.~\ref{fig1}), $\Omega_{\Lambda}^0 = +0.35$, $\nu =
-0.2$ and $w_X=-0.6$. In this case, $P_{\Lambda \rm{X}}(k)$ presents
a slight deviation as compared to $P_{\Lambda}(k)$ {at large scales
(i.e. at small $k$)}. {The lower set of curves in (b) displays the
real (unnormalized) growth, see the text}. } \label{fig4}
\end{figure}

The $\CC$XCDM power spectrum was calculated for two fiducial values
of the DE speed of sound, $c_{s}^2=1$ and $c_{s}^2=0.1$ and several
combinations of the parameters $\nu$, $\wX$ and $\OLo$. For values
of the parameters allowed in Figs.~\ref{fig1}-\ref{fig2} (shaded and
striped region) we find that $A_{\CC\rm X}\approx 1$ (within $\sim
10\%$ of accuracy).


In Fig.~\ref{fig4}a, we put together the 2dFGRS observed galaxy
power spectrum, the $\CC$CDM spectrum and the normalized and
unnormalized $\CC$XCDM one, for the set of parameters
$\Omega_{\Lambda}^0 = 0.8$, $\nu = \nu_0\equiv 2.6\times 10^{-2} $
and $w_X=-1.6$, which are allowed in Figs.~\ref{fig1}-\ref{fig2}.
For these values, we have obtained a normalization factor $A_{\CC
X}\cong 1.1$ and an accurate agreement between the ${\CC\rm XCDM}$
power spectrum and $P_{\CC}(k)$. This was expected since we are
assuming allowed values of the parameters, i.e., values already
consistent with LSS data according to the `effective' approach used
in \cite{GOPS} (cf. the discussion in section
\ref{sect:perturbLXCDM}). Therefore, the predicted power spectrum
from the $\CC$XCDM ought to be very close to the ${\CC\rm CDM}$ one,
which is in fact what we have substantiated now by explicit
numerical check.

However, for values of the parameters out of the allowed region in
Fig.~\ref{fig1} the predicted matter power spectrum can differ
significantly from the $\CC$CDM one, $P_{\CC}(k)$. This occurs
mainly for points that do not satisfy the ``F-test" condition
\cite{GOPS}, even if the other observable constraints (namely the
ones related to nucleosynthesis and the present value of the EOS
(cf. section \ref{sect:perturbLXCDM})) are fulfilled. Let us remind
that the F-test consists in requiring that the matter power spectrum
of the model under consideration (in this case, the $\CC$XCDM model)
differs from that of the $\CC$CDM in less that a 10\%, under the
assumption that DE perturbations can be neglected. Given the fact
(explicitly analyzed here) that the DE perturbations should not play
a very important role, it is reasonable to expect that the F-test
should be approximately valid even when we do not neglect the DE
perturbations. Thus, the $\CC$XCDM model should exhibit a large
deviation in the amount of growth with respect to the $\CC$CDM
precisely for those points failing the F-test. Points of this sort
are those located in the striped region, but outside the shaded one
in Fig.~\ref{fig1}. For these points, we should expect an
anomalously large normalization factor $A_{\CC X}$ (namely, the
factor that controls the matching of the two overall shapes) and, at
the same time, we may also observe an evident scale dependence in
the power spectrum, i.e. some significant difference in the
predicted shape as compared to the $\CC$CDM one. Such potentially
relevant scale dependence (or $k$-dependence) is introduced by the
DE perturbations themselves through the last term on the
\textit{r.h.s} of (\ref{pnadDE}) and is eventually fed into
equations (\ref{DEPert})-(\ref{PertEeqDE}).

As a concrete example, let us consider Fig.~\ref{fig4}b where we
compare the 2dFGRS observed galaxy power spectrum and $P_{\CC}(k)$
with the $\Lambda$XCDM matter power spectrum $P_{\CC X}(k)$ for the
following set of parameters: $\OLo = +0.35$, $\nu = -0.2$ and
$w_X=-0.6$. These values fulfill the nucleosynthesis bound
(constraint No. 1 in section \ref{sect:perturbLXCDM}), specifically,
we have $|\epsilon|=0.08$ for these parameters (meaning that DE
density at the nucleosynthesis time represents roughly only $8\%$ of
the total energy density); and satisfy also the current EOS
constraint (No. 3 in section \ref{sect:perturbLXCDM}): $\we^0=-0.8$.
However, this choice of parameters largely fails to satisfy the
constraint No. 4, i.e. the F-test: indeed, we find $F=2.06$, which
implies that the discrepancy in the amount of growth with respect to
the $\CC$CDM when we neglect DE perturbations is more than $200\%$!
As expected, for such set of parameters we encounter a large
normalization factor for the two fiducial DE sound speeds
$c_{s}^2=1$ and $c_{s}^2=0.1$ that we are using in our analysis (on
average $A_{\CC X}\simeq 2.7$). This is reflected in the evident gap
existing between the upper and lower set of curves in
Fig.~\ref{fig4}b. The lower set reflects the real growth
$|\delta_M(k)|^2$ of matter perturbations before applying the
normalization factor. Such normalization consists in the following:
for the smallest scale available in the data, the $\CC$XCDM curves
have been shifted upwards until they match up with the standard
$\CC$CDM prediction. Apart from the overall gap between the two set
of curves, we also find a significant shape deviation with respect
to the standard $\CC$CDM model at large scales, as it is patent in
Fig.~\ref{fig4}b. This feature is more clearly seen at small sound
speeds, see next section.


\section{Matter and dark energy density
fluctuations}\label{DEfluct}

As we have discussed in section \ref{sect:genericfeatures}, the DE
fluctuations $\de_D$ should oscillate and become eventually
negligible as compared to the matter fluctuations $\de_M$, specially
at small scales (inside the sound horizon). However, as also noted
above, for values that significantly violate the
F-test\,\cite{GOPS}, the power spectrum and its shape can be
noticeably different from that of the $\CC$CDM model (cf.
Fig.~\ref{fig4}b). This suggests that, under such circumstances, the
DE density perturbations are not completely negligible owing to the
fact that the term which depends on $k$ in the perturbation
equations is also proportional to $\de_D$ -- see
Eq.\,(\ref{DEPertTheta}). In addition, there appears a suppression
of the growth of matter fluctuations in comparison with the growth
predicted by the $\Lambda$CDM model. This inhibition of matter
growth is characteristic of cosmologies where the DE behaves
quintessence-like, i.e. when the DE density decreases with the
expansion, whereas phantom-like DE (increasing with the expansion)
would cause the opposite effect (an enhancement of the power). A
similar situation was also observed in Ref.\cite{FabrisDens} for
models with pure running $\CC$, where in the case $\nu>0$ (in which
$\CC$ decreases with the expansion) there is an inhibition of growth
while for $\nu<0$ (when $\CC$ increases with the expansion) there is
an enhancement  -- see also \cite{Abramo07,BCFP08} and \cite{HWA08}
for other studies.


\begin{figure}[t]
\resizebox{\columnwidth}{!}{\includegraphics{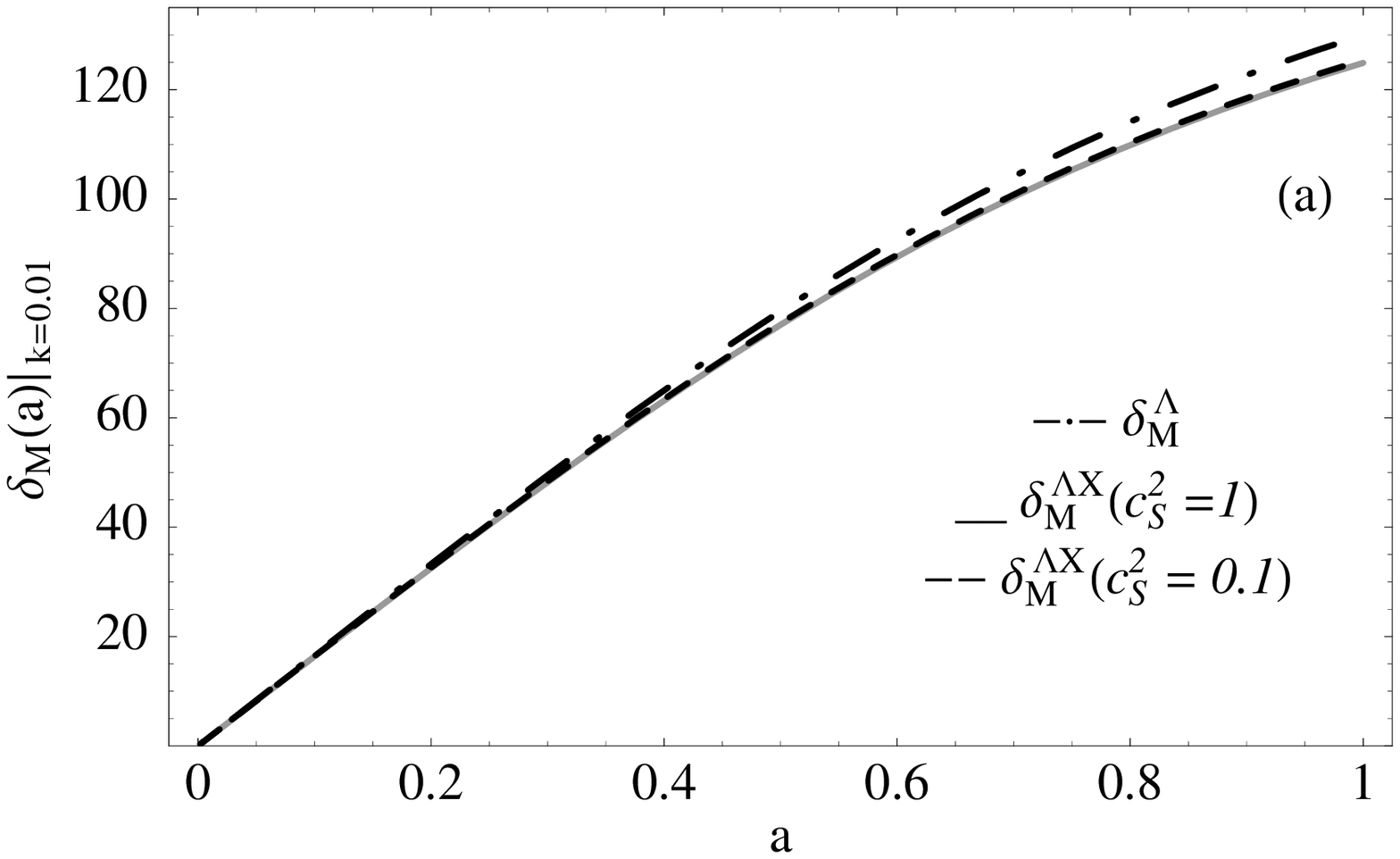}}\\[3ex]
\resizebox{\columnwidth}{!}{\includegraphics{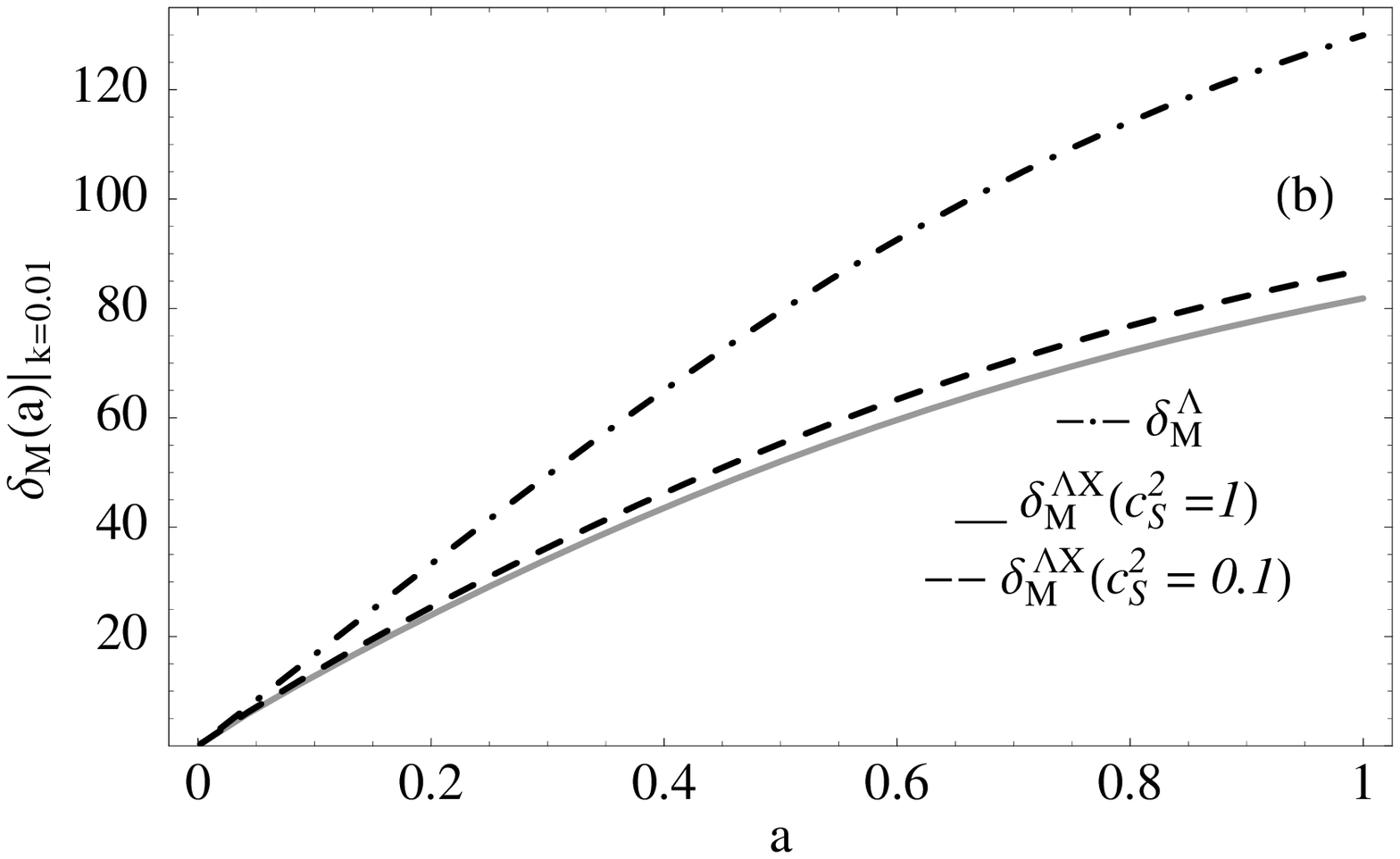}}
\caption{The $\Lambda$XCDM matter density fluctuations at a fixed large scale
$k=0.01$ (in units of $h\,$Mpc$^{-1}$) as a function of the scale
factor $a$ in comparison with those predicted by the $\Lambda$CDM
model (dot-dashed line). For the former we have assumed the same
values of the parameters and meaning of the lines as in
Fig.~\ref{fig4}: (a) for the set of parameters in the allowed
region; (b) for the set of parameters not allowed by the F-test in
\cite{GOPS}.} \label{fig5}
\end{figure}



\begin{figure}
\resizebox{\columnwidth}{!}{\includegraphics{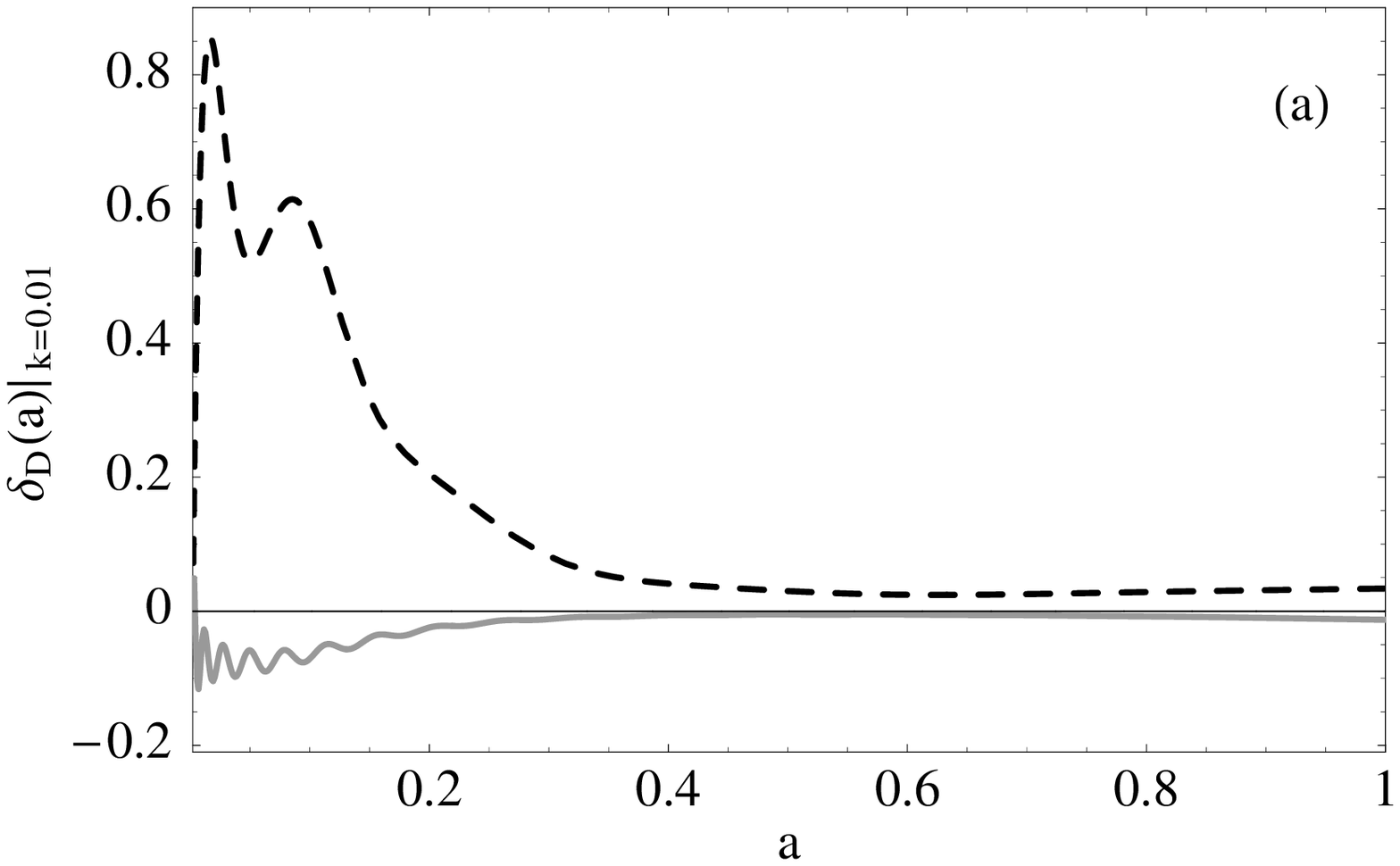}}\\[3ex]
\resizebox{\columnwidth}{!}{\includegraphics{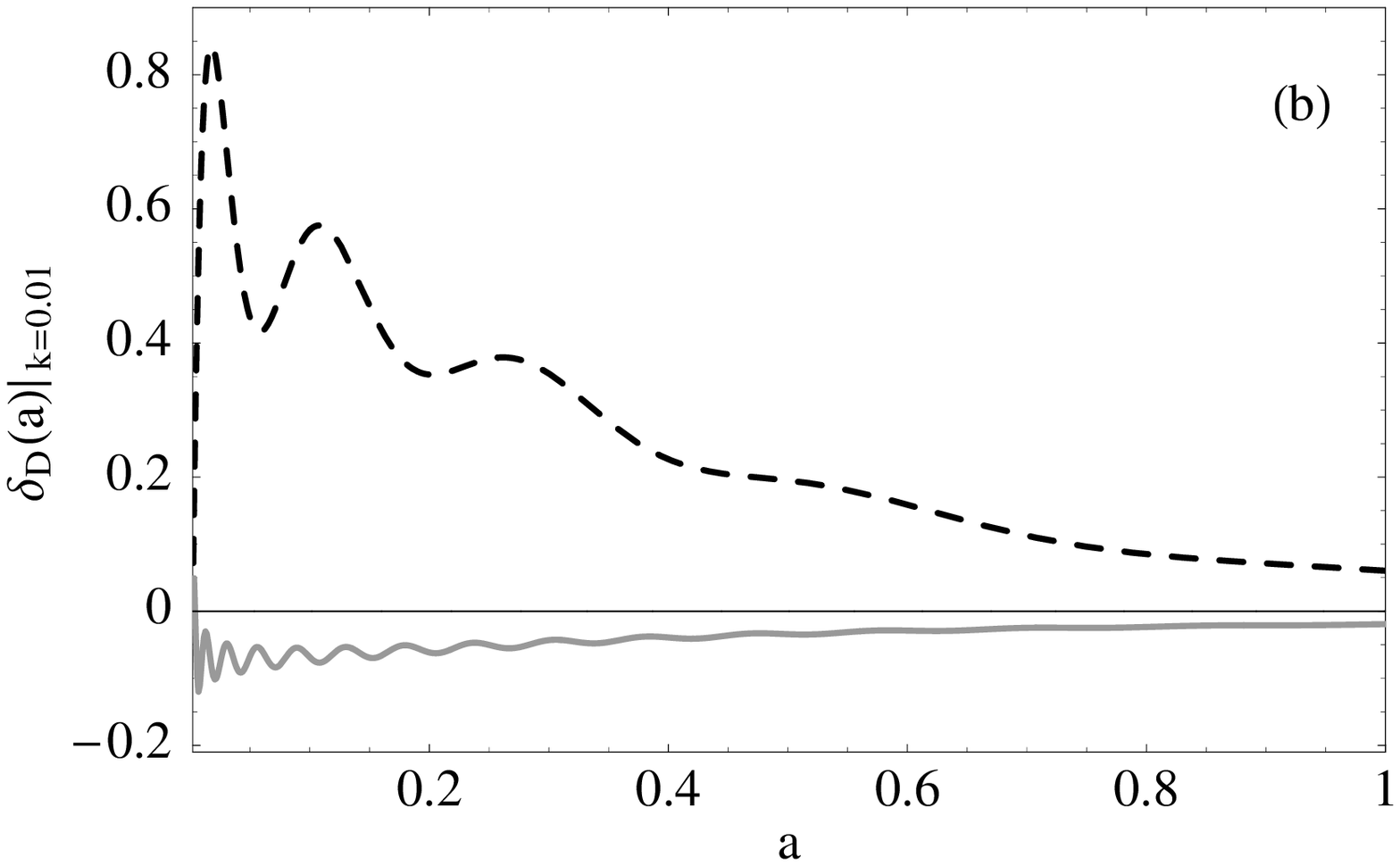}}
\caption{The $\CC$XCDM growth of DE fluctuations at the large scale $k=0.01$ (in
units of $h\,$Mpc$^{-1}$) as a function of the scale factor $a$. We
have assumed the same values of the parameters as in Fig.~\ref{fig4}
for DE sound speed $c_{s}^2=0.1$ (dashed line) and $c_{s}^2=1$
(solid/gray line): (a) for the set of parameters in the allowed
region; (b) for the set of parameters not allowed by the F-test in
\cite{GOPS}. } \label{fig6}
\end{figure}


{Let us clarify that these differences in the amount of growth are
present even if we neglect the DE perturbations, see \cite{GOPS}. In
fact, the effect of the latter is very small, specially for allowed
values of the parameters, and becomes noticeable only at large
scales. At these scales, we find that the DE perturbations tend to
compensate the suppression produced at the background level. This
slight enhancement is greater the smaller is the DE sound speed}.
Such feature can be appreciated in Fig.~\ref{fig5}, where we compare
the growth of the matter fluctuations at a large scale $\ell\sim
k^{-1}$ (with $k=0.01$) predicted by both the $\Lambda$CDM and
$\Lambda$XCDM models for the two fiducial sound speeds of the DE
considered before and for the same values of the parameters as in
Fig.~\ref{fig4}.

The growth of matter density fluctuations for the $\Lambda$XCDM
model is in agreement with the predicted one by the $\Lambda$CDM
model (the dot-dashed and black line) in Fig.~\ref{fig5}a, for the
set of parameters in the allowed region, whereas in Fig.~\ref{fig5}b
we see the previously commented suppression for the set of
parameters not satisfying the F-test. The former case can be
compared with Fig.~\ref{fig6}a in which we have assumed the same set
of allowed parameters; as expected, we find completely negligible DE
fluctuations today {and in the recent past}, in agreement with the
F-test assumption \cite{GOPS} which means completely negligible DE
fluctuations at large scales and a maximum $10\%$ of deviation from
the $\Lambda$CDM growth of matter density fluctuations. On the other
hand, values of the parameters not satisfying the F-test present not
only suppression on the growth of matter density fluctuations, as
shown in Fig.~\ref{fig5}b, but also larger DE fluctuations today
{and in the recent past}, as shown in Fig.~\ref{fig6}b.

\begin{figure}[t]
\begin{center}
\resizebox{\columnwidth}{!}{\includegraphics{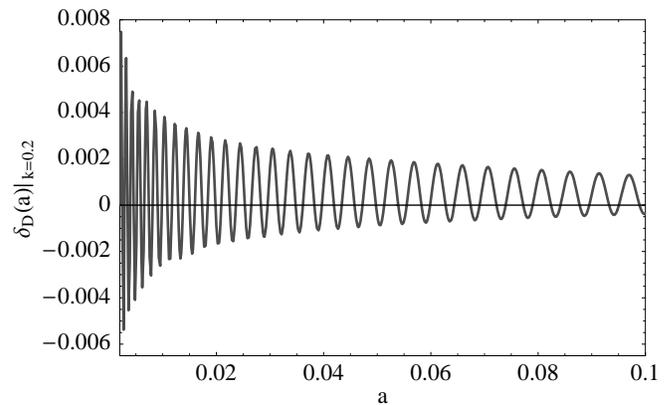}}
\caption{The $\CC$XCDM growth of DE fluctuations for a small scale $k=0.2$ (in
units of $h\,$Mpc$^{-1}$) and the same set of allowed parameters
assumed before, and for DE sound speed $c_{s}^2=0.1$.} \label{fig7}
\end{center}
\end{figure}

Furthermore, as discussed in Section \ref{sect:genericfeatures}, the
growth of DE fluctuations is expected to oscillate at small scales
and rapidly decay, what legitimate our assumptions for the initial
conditions of the DE perturbations. We show these oscillations for
an allowed set of parameters in Fig.~\ref{fig7}. Similar behavior is
obtained for values of the parameters not allowed by the F-test and
for both DE sound speeds $c_{s}^2=1$ and $c_{s}^2=0.1$. The
amplitude of the DE growth starts negligible ($\sim 10^{-3}$) and
rapidly decay to zero, as shown in Fig.~\ref{fig7}.


\begin{figure}
\resizebox{\columnwidth}{!}{\includegraphics{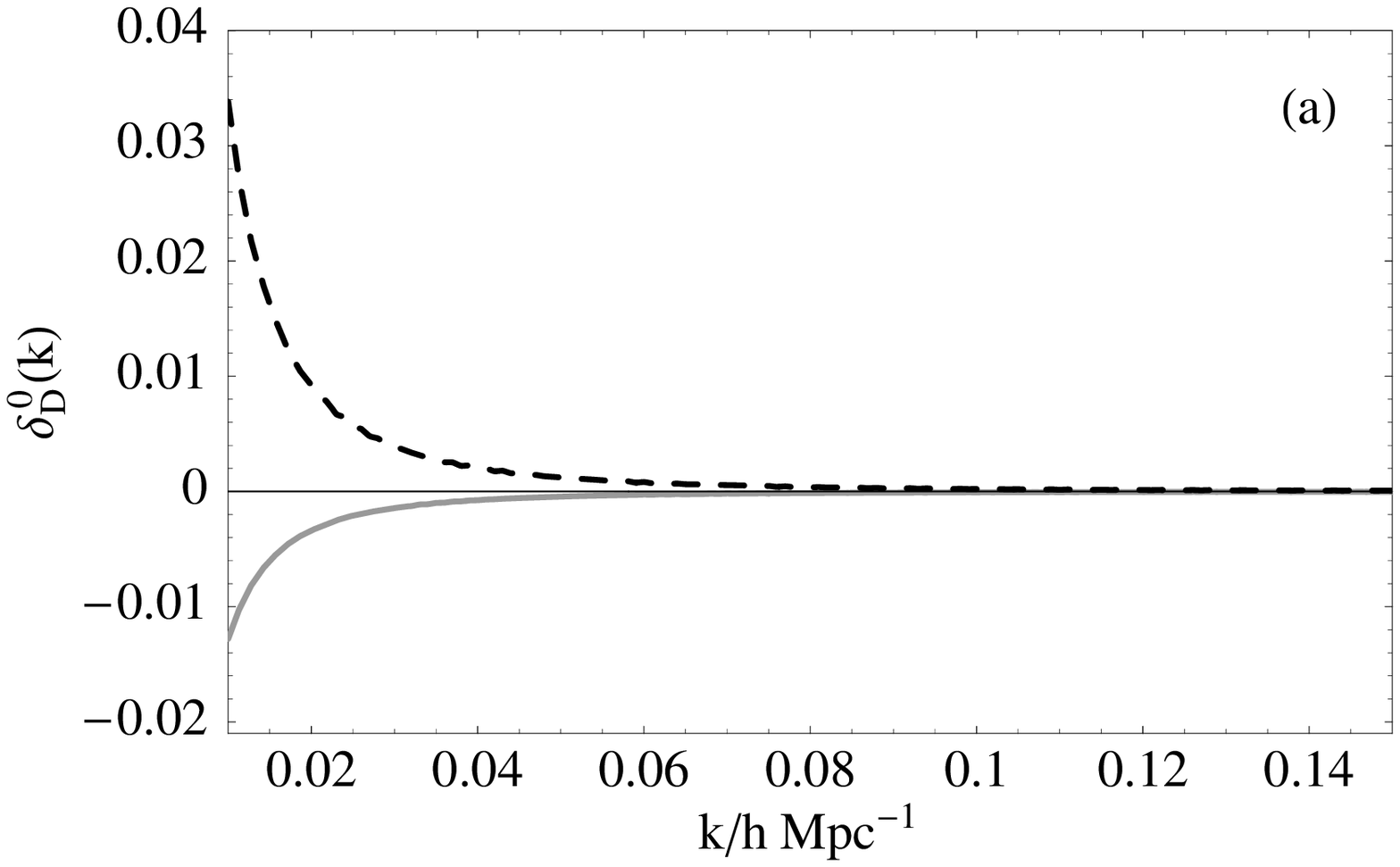}}\\[3ex]
\resizebox{\columnwidth}{!}{\includegraphics{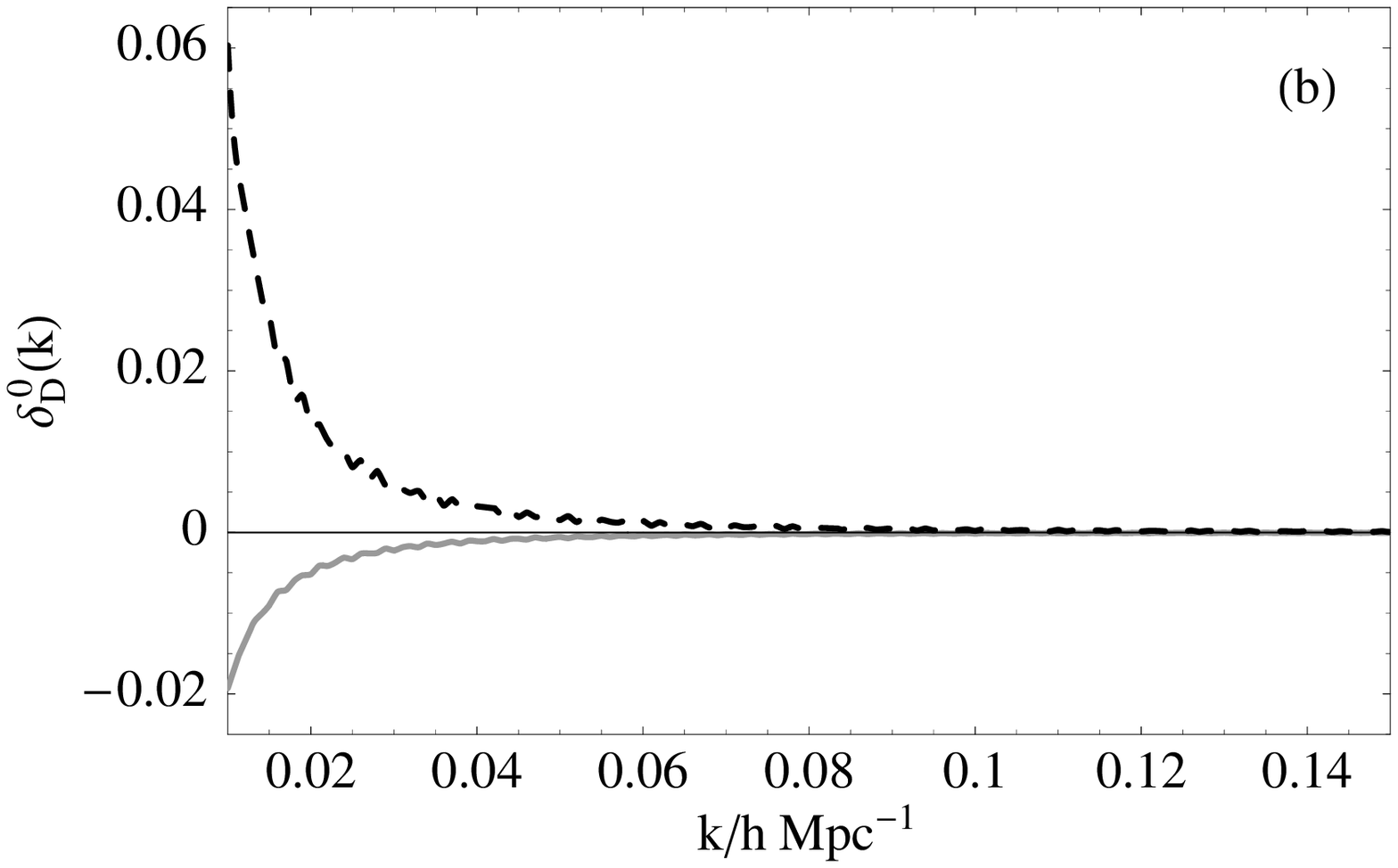}}
\caption{The  scale dependence of DE fluctuations today ($a=1$) for the same values of
parameters and meaning of the lines as in Fig.~\ref{fig6}. They
rapidly decay at small scales (large $k$). } \label{fig8}
\end{figure}

In Fig.~\ref{fig8} we plot the present value of the DE perturbations
as a function of the wave number for the two sets of allowed
(Fig.~\ref{fig8}a) and non-allowed (Fig.~\ref{fig8}b) parameters
used in the previous plots. We see that the DE perturbations are
negligible at small scales (large $k$), whereas they become larger
at larger scales. This is because by increasing the scale we are
getting closer to the sound horizon, as discussed in
section~\ref{sect:genericfeatures}. We also see that the DE
perturbations are larger for the parameters not allowed by the
F-test, which explains why the shape of the matter power spectrum
differs from that of the $\CC$CDM in this case  (cf.
Fig.~\ref{fig4}b). However, when comparing with Fig.~\ref{fig5}, we
see that even at the largest explored scale ($k=0.01$) the ratio
$\de_D/\de_M$ remains rather small, staying at the level of
$10^{-3}$. Finally, let us comment that the DE density contrast can
become negative with the evolution, as in this case happens for
$\csd=1$.


\begin{figure}[t]
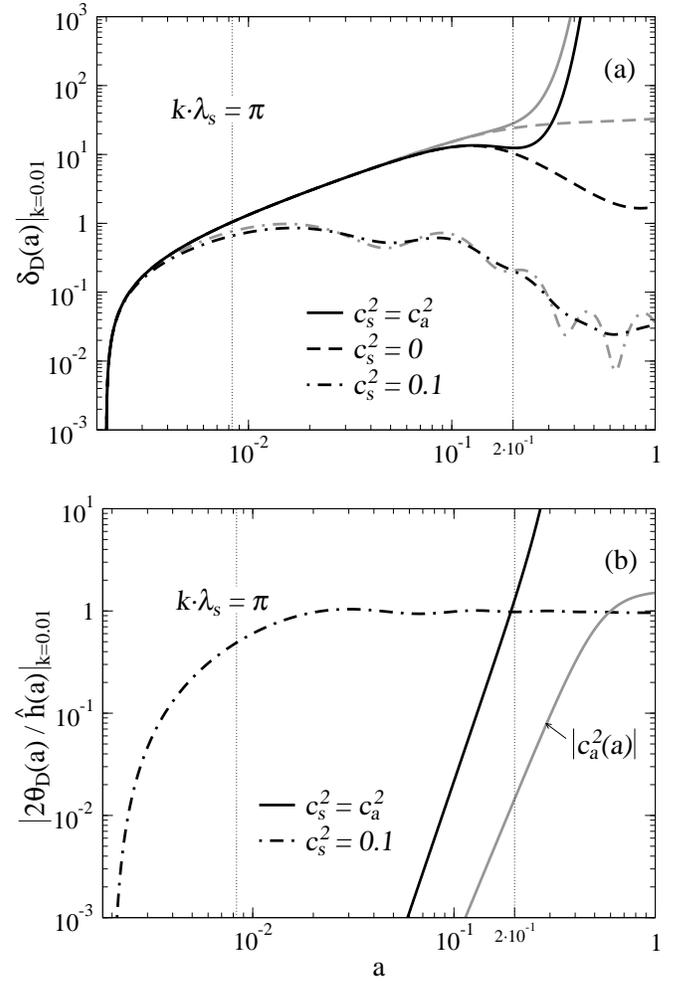

\center
\resizebox{\columnwidth}{!}{\includegraphics{Figures/fig8a.eps}}
\resizebox{\columnwidth}{!}{\includegraphics{Figures/fig8b.eps}}
\caption{(a) {Evolution of the DE perturbations for the same set of
parameters as in Fig.\,\ref{fig4}a, at the large scale $k=0.01$ and
for three different speeds of sound: $\csd=\cad<0$ (solid line),
$\csd=0$ (dashed line) and $\csd=0.1$ (dot-dashed line). For the
latter, the DE perturbations begin to decay after the sound horizon
crossing (characterized by the condition $k\lambda_s=\pi$), whereas
in the adiabatic case  $\delta_D$ starts to grow exponentially at
$a\simeq0.2$. The last term in Eq.~(\ref{DEPert}) may be neglected
(gray lines) without altering the qualitative behavior, which is
triggered by the $\theta_D$-term and, ultimately, by the one
proportional to $k^2$ in (\ref{DEPertTheta}); (b) Comparison of the
two terms inside the curly brackets in (\ref{DEPert}) for the
$\csd=0.1$ and adiabatic cases. When the $\theta_D$-term becomes
important, the stabilization (resp. unbounded growth) of the
non-adiabatic (resp. adiabatic) perturbations becomes manifest.}}
\label{fig9}
\end{figure}


{As discussed in section~\ref{sect:genericfeatures}, the decay of
the DE perturbations takes place once the term proportional to $k^2$
in (\ref{DEPertTheta}) becomes dominant. This same term is also
responsible for the exponential growth of the (DE) perturbations in
the adiabatic case or, more generally, whenever $\csd$ is negative.
In order to better appreciate its influence, it is useful to compare
the evolution of the DE perturbations in the adiabatic case (for the
most common situation where $c_a^2<0$) and the non-adiabatic one
($\csd>0$) with the scenario in which $\csd=0$, since in the latter
the term proportional to $k^2$ disappears from the equations.} {This
is precisely what has been done in Fig.~\ref{fig9}a for the allowed
set of parameters used throughout this section. In that figure, it
is shown the evolution of the DE density contrast at a sufficiently
large scale $\ell\sim k^{-1}$ for which the DE perturbations can be
sizeable (namely at $k=0.01$). We illustrate the effect for three
different regimes of the speed of sound: $\csd=\cad$ (with
$\cad<0$), $\csd=0$ and $\csd=0.1$. The gray lines represent the
evolution of $\delta_D$ when the last term in Eq.(\ref{DEPert}) is
neglected, showing indeed that the qualitative behavior of the
perturbations in the adiabatic and $\csd=0.1$ cases does not stem
from that term.}

For $\csd=0.1$, the scale considered is initially (i.e. at
$a_i=1/500$) larger than the sound horizon, and thus the term
proportional to $k^2$ is negligible at the beginning of the
evolution. The same is true for the adiabatic case because, in the
asymptotic past, the effective EOS of the $\CC$XCDM model resembles
that of matter-radiation ($\we(a)\to w_m$ for $a\to
0$)\,\cite{LXCDM12} and, thus, we have $\csd=\cad\simeq\we\simeq0$
in the matter dominated epoch. (We recall that in all our discussion
we remain in the matter epoch, equality being at $a\sim 10^{-4}$).
Therefore, the term proportional to $k^2$ is initially unimportant
in all the three cases and this makes the perturbations to evolve in
a nearly identical fashion at these first stages, as it can be
clearly seen from Fig.~\ref{fig9}a.

{As the evolution continues, the curve corresponding to $\csd=0.1$
begin to depart from the others. This occurs mainly from the instant
when the sound horizon is crossed, i.e. when the wavelength of the
$k$-mode gets comparable to the sound horizon; such instant can be
defined through the condition $k\lambda_s=\pi$, similarly as in
\cite{Hu98}. Then, the term proportional to $k^2$ begins to
dominate, which in turn makes the DE velocity gradient $\theta_D$ to
rapidly increase and the DE perturbations to decay. Later on, the
term proportional to $k^2$ becomes important also in the adiabatic
case. Due to the different sign ($c_a^2<0$), the effect that it
triggers is now opposite to the one observed in the $\csd=0.1$ case:
thus, instead of getting stabilized, the DE perturbations initiate
an exponential growth.}

{The previous features can be further assessed in a quantitative way
by comparing the numerical importance (in absolute value) of the two
terms inside the curly brackets in (\ref{DEPert}). In
Fig.~\ref{fig9}b, we plot the ratio between these two terms for both
the adiabatic case (with $c_a^2<0$) and the non-adiabatic situation
($\csd=0.1$) (note that the term proportional to $H^2/k^2$ may be
neglected for the sub-Hubble perturbations we are dealing with).
Comparison with Fig.~\ref{fig9}a reveals that it is precisely when
the term proportional to $\theta_D$ stops being negligible that the
evolution of the perturbations begins to depart from the $\csd=0$
case. The absolute value of the adiabatic speed of sound is also
shown, in order to illustrate the ultimate reason for the start-up
of the exponential growth in the adiabatic mode: it is only when
$\cad$ begins to depart significantly from 0 that the term
proportional to $k^2$ becomes important, which in turn triggers a
rocket increase of the velocity gradient $\theta_D$.}

{As we have discussed in connection to Fig.~\ref{fig9}a, the initial
evolution of the perturbations in the $\CC$XCDM model is nearly the
same for any of the three values of the speed of sound. In fact, in
the adiabatic case, and given the behavior of the effective EOS in
the asymptotic past, the conditions that lead to the simplified
setup (\ref{DEPert2}) hold. Therefore, that is the equation
initially controlling the evolution of $\delta_D$. For $\csd=0$ we
arrive at exactly the same equation, whereas for positive $\csd$ the
resulting equation only differs by the last term in (\ref{DEPert}),
which, as it has been previously discussed, happens to be negligible
at least during the first stages of the evolution. Notice that for
$\we\simeq const.$, Eq.~(\ref{DEPert2}) integrates to
$\delta_D(a)=(1+\we)\,\delta_M+ C$, where $C$ is a constant
determined by the initial conditions. In section \ref{sect:effEOS},
we pointed out that $C=0$ corresponds to the adiabatic initial
condition (\ref{adiabcond}). However, for the alternate initial
condition (\ref{DEic}), and taking into account that $\we(a_i)\simeq
w_M=0$ for the $\CC$XCDM model in the early matter dominated
epoch\,\cite{LXCDM12}, we have $C=- \delta_M(a_i)$ and thus
$\delta_D(a)=\,\delta_M(a) -\delta_M(a_i)$. From here we find that
the ratio between DE and matter perturbations in the early times of
the evolution reads:}
\begin{equation}\label{ratsim}
\frac{\delta_D(a)}{\delta_M(a)}=1
-\frac{\delta_M(a_i)}{\delta_M(a)}\,.
\end{equation}

\begin{figure}[t]
\begin{center}
\resizebox{\columnwidth}{!}{\includegraphics{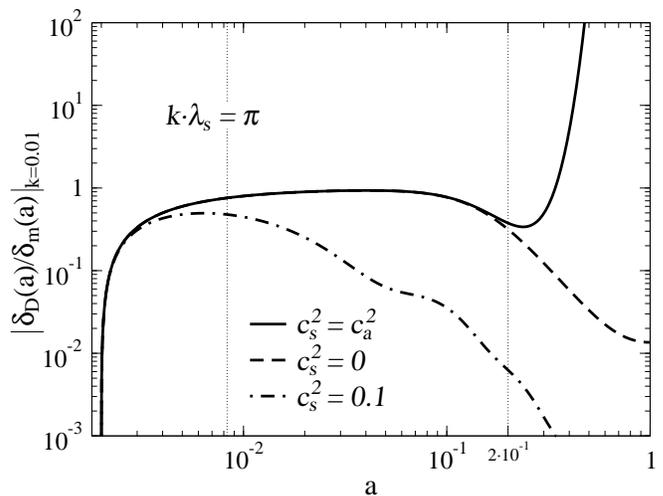}}
\caption{{Evolution of the ratio between DE and matter perturbations
for the same parameter values as in Fig.~\ref{fig9}. The special
regime of perturbations (\ref{DEPert2}) is seen to be approximately
realized during the first stages of the evolution, for any of the
three considered speeds of sound. Since $\we\to 0$ for small $a$ in
the $\CC$XCDM model, we expect $\delta_D/\delta_M\simeq 1$ (cf.
Eq.~(\ref{ratsim})) until the conditions leading to (\ref{DEPert2})
no longer hold. One can clearly confirm this situation in the
figure.}} \label{fig10}
\end{center}
\end{figure}

{This simple predicted behavior is confirmed from the numerical
analysis in Fig.~\ref{fig10}, where again the allowed set of
parameters has been used. We see that the ratio $\delta_D/\delta_M$
starts being 0, and subsequently as the matter perturbations grow
the last term in (\ref{ratsim}) diminishes, until the asymptotic
value $\delta_D/\delta_M=1$ is reached. This value is maintained
until the conditions leading to (\ref{DEPert2}) cease to be valid.
In the adiabatic and $\csd>0$ cases, this happens when the term
proportional to $k^2$ on the \textit{r.h.s.} of
Eq.\,(\ref{DEPertTheta}) can no longer be neglected. On the other
hand, when $\csd=0$, the $\delta_D/\delta_M\simeq 1$ regime is
abandoned at the point when the effective EOS starts acquiring
sizable negative values. Moreover, being the term proportional to
$k^2$ absent, it is now the last term on the \textit{r.h.s.} of
Eq.\, (\ref{DEPert}) -- which was irrelevant for the other two cases
-- the one that tends to stabilize the DE perturbations (see also
Fig.~\ref{fig9}a).}


From the detailed analysis that we have presented here, we conclude
that the approximation of neglecting the DE perturbations can be
justified\,\cite{GOPS}. But this does not mean that the computation
of these perturbations is useless. Indeed, the issue at stake here
is not so much the quantitative impact of the DE fluctuations upon
the matter power spectrum -- which is actually negligible, as we
have seen -- but rather the fact that the DE perturbations may be
consistently defined in a certain subregion of the parameter space
only. Of course this subregion cannot be detected within the context
of the effective approach \,\cite{GOPS}. Therefore, in general, the
computation of the DE perturbations may have a final quantitative
bearing on this kind of analyses since it may further restrict the
physical region of the parameter space in a very significant way. In
short, even though the simultaneous account of the DE perturbations
has a small numerical effect on the matter power spectrum within the
domain where the full system of cosmological perturbations is
well-defined, it may nevertheless prove to be a highly efficient
method for excluding large regions of parameter space where that
system is ill-defined. The upshot is that the combined analysis of
the DE and matter perturbations may significantly enhance the
predictivity of the model, as we have indeed illustrated in detail
for the non-trivial case of the $\CC$XCDM model of the cosmic
evolution.

\noindent \section{Discussion and conclusions}
\label{sect:conclusions}
In this paper, we have addressed the impact of the cosmological
perturbations on the coincidence problem. In contrast to the
previous study\,\cite{GOPS}, where this problem was examined in a
simplified ``effective approach'' in which the dark energy (DE)
perturbations were neglected, in the present work we have taken them
into account in a full-fledged manner. We find that the results of
the previous analysis were reasonable because the DE perturbations
generally tend to smooth at scales below the sound horizon. However,
the inclusion of the DE perturbations proved extremely useful to pin
down the physical region of the parameter space and also to put the
effective approach within a much larger perspective and to set its
limitations.

First of all, we have performed a thorough discussion on the coupled
set of matter and DE perturbations for a general multicomponent
fluid. This has prepared the ground to treat models in which the DE
is a composite medium with a variable equation of state (EOS). We
have concentrated on those cases in which the DE, despite its
composite nature, is described by a self-conserved density $\rD$.
Notice that if matter is covariantly conserved, the covariant
conservation of the DE is mandatory. In particular, this is the
situation for the standard $\CC$CDM model, although in this case the
self-conservation of the DE appears through a trivial cosmological
constant term, $\rL=\rL^0$, which remains imperturbable throughout
the entire history of the Universe. One may nevertheless entertain
generalized frameworks where the DE is not only self-conserved, but
is non-trivial and dynamical. This is not a mere academic exercise;
for instance, in quantum field theory in curved space-time we
generally expect that the vacuum energy should be a running
quantity\,\cite{JHEPCC1,SSIRGAC06,ShS08}. Therefore, in such cases,
the CC density becomes an effective parameter that may evolve
typically with the expansion rate, $\rL=\rL(H)$, and constitutes a
part of the full (dynamical) DE of the composite cosmological system
with variable EOS. In these circumstances, if the gravitational
coupling $G$ is constant, the running CC density $\rL=\rL(H)$ cannot
be covariantly conserved unless other terms in the effective action
of this system compensate for the CC variation. We have called the
effective entity that produces such compensation ``$X$'' or
``cosmon'', and denoted with $\rX$ its energy density. Therefore,
$\rD=\rL+\rX$ is the self-conserved total DE density in this
context, which must be dealt with together with the ordinary density
of matter $\rM$. A generic model of this kind is what we have called
the $\CC$XCDM model\,\cite{LXCDM12,GOPS}.

Furthermore, from general considerations  based on the covariance of
the effective action of QFT in curved
space-time\,\cite{JHEPCC1,SSIRGAC06,ShS08}, we expect that the
running CC density $\rL=\rL(H)$ should be an affine quadratic law of
the expansion rate $H$, see Eq.\,(\ref{runlamb}). Using this guiding
principle and the ansatz of self-conservation of the DE, we find
that the evolution of $\rX$, and hence of $\rD$, becomes completely
determined, even though its ultimate nature remains unknown. In
particular, $X$ is \textit{not} a scalar field in general.

The $\CC$XCDM model was first studied in \cite{LXCDM12} as a
promising solution to the cosmic coincidence problem, in the sense
that the coincidence ratio $r=\rD/\rM$ can stay relatively constant,
meaning that it does not vary in more than one order of magnitude
for many Hubble times. The main aim of the present paper was to make
a further step to consolidate such possible solution of the
coincidence problem, specifically from the analysis of the coupled
system of matter and DE perturbations. Let us remark that this has
been a rather non-trivial test for the $\CC$XCDM model. Indeed,
after intersecting the region where the DE perturbations of this
model can be consistently defined, with the region where the
coincidence problem can be solved\,\cite{LXCDM12,GOPS}, we end up
with a significantly more reduced domain of parameter space where
the model can exist in full compatibility with all known
cosmological data. The main conclusion of this study is that the
predictivity of the model has substantially increased. Therefore, it
can be better put to the test in the next generation of precision
cosmological observations, which include the promising DES, SNAP and
PLANCK projects\,\cite{SNAP}.

Interestingly enough, we have found that the final region of the
parameter space is a naturalness region which is more accessible to
the aforementioned precision experiments. For example, we have
obtained the bound $0\leqslant\nu\lesssim \nu_0\sim 10^{-2}$ for the
parameter that determines the running of the cosmological term. This
bound is perfectly compatible with the physical interpretation of
$\nu$ from its definition (\ref{nu}). Moreover, our analysis
indicates that the cosmon entity $X$ behaves as ``phantom
matter''\cite{LXCDM12}, i.e. it satisfies $\wX<-1$ with negative
energy density. This result is a clear symptom (actually an expected
one) from its effective nature. It is also a welcome feature; let us
recall\,\cite{LXCDM12} that ``phantom matter'', in contrast to the
``standard'' phantom energy, prevents the Universe from reaching the
Big Rip singularity. Finally, perhaps the most noticeable (and
experimentally accessible) feature that we have uncovered from the
analysis of the DE perturbations in the $\CC$XCDM model, is that the
overall EOS parameter $\we$ associated to the total DE density $\rD$
behaves effectively as quintessence ($\we\gtrsim -1$) in precisely
the region of parameter space where the cosmic coincidence problem
can be solved. In other words, quintessence is mimicked by the
$\CC$XCDM model in that relevant region, despite that there is no
fundamental quintessence field in the present framework. A detailed
confrontation of the various predictions of the $\CC$XCDM model (in
particular, the kind of dependence $\we=\we(z)$) with the future
accurate experimental data\,\cite{SNAP}, may eventually reveal these
features and even allow to distinguish this model from alternative
DE proposals based on fundamental quintessence fields.

To summarize: we have demonstrated that the set of cosmological
models characterized by a composite, and covariantly conserved, DE
density $\rD$ in which the vacuum energy $\rL$ is a dynamical
component (specifically, one that evolves quadratically with the
expansion rate, see Eq.\,(\ref{runlamb})), proves to be a
distinguished class of models that may provide a consistent
explanation of why $\rD$ is near $\rM$, in full compatibility with
the theory of cosmological perturbations and the rest of the
cosmological data. Remarkably, such class of models is suggested by
the above mentioned renormalization group approach to cosmology. We
conclude that the $\CC$XCDM model can be looked upon as a rather
predictive framework that may offer a robust, and theoretically
motivated, dynamical solution to the cosmic coincidence problem.

\vspace{1cm}

{\bf Acknowledgments.}\ The authors are grateful to Julio Fabris for
useful discussions. We have been supported in part by MEC and FEDER
under project FPA2007-66665 and also by DURSI Generalitat de
Catalunya under project 2005SGR00564. The work of J.G. is also
financed by MEC under BES-2005-7803. We acknowledge the support from
the Spanish Consolider-Ingenio 2010 program CPAN CSD2007-00042.

\newcommand{\JHEP}[3]{ {JHEP} {#1} (#2)  {#3}}
\newcommand{\NPB}[3]{{\sl Nucl. Phys. } {\bf B#1} (#2)  {#3}}
\newcommand{\NPPS}[3]{{\sl Nucl. Phys. Proc. Supp. } {\bf #1} (#2)  {#3}}
\newcommand{\PRD}[3]{{\sl Phys. Rev. } {\bf D#1} (#2)   {#3}}
\newcommand{\PLB}[3]{{\sl Phys. Lett. } {\bf B#1} (#2)  {#3}}
\newcommand{\EPJ}[3]{{\sl Eur. Phys. J } {\bf C#1} (#2)  {#3}}
\newcommand{\PR}[3]{{\sl Phys. Rep. } {\bf #1} (#2)  {#3}}
\newcommand{\RMP}[3]{{\sl Rev. Mod. Phys. } {\bf #1} (#2)  {#3}}
\newcommand{\IJMP}[3]{{\sl Int. J. of Mod. Phys. } {\bf #1} (#2)  {#3}}
\newcommand{\PRL}[3]{{\sl Phys. Rev. Lett. } {\bf #1} (#2) {#3}}
\newcommand{\ZFP}[3]{{\sl Zeitsch. f. Physik } {\bf C#1} (#2)  {#3}}
\newcommand{\MPLA}[3]{{\sl Mod. Phys. Lett. } {\bf A#1} (#2) {#3}}
\newcommand{\CQG}[3]{{\sl Class. Quant. Grav. } {\bf #1} (#2) {#3}}
\newcommand{\JCAP}[3]{{ JCAP} {\bf#1} (#2)  {#3}}
\newcommand{\APJ}[3]{{\sl Astrophys. J. } {\bf #1} (#2)  {#3}}
\newcommand{\AMJ}[3]{{\sl Astronom. J. } {\bf #1} (#2)  {#3}}
\newcommand{\APP}[3]{{\sl Astropart. Phys. } {\bf #1} (#2)  {#3}}
\newcommand{\AAP}[3]{{\sl Astron. Astrophys. } {\bf #1} (#2)  {#3}}
\newcommand{\MNRAS}[3]{{\sl Mon. Not. Roy. Astron. Soc.} {\bf #1} (#2)  {#3}}
\newcommand{\JPA}[3]{{\sl J. Phys. A: Math. Theor.} {\bf #1} (#2)  {#3}}
\newcommand{\ProgS}[3]{{\sl Prog. Theor. Phys. Supp.} {\bf #1} (#2)  {#3}}
\newcommand{\APJS}[3]{{\sl Astrophys. J. Suppl.} {\bf #1} (#2)  {#3}}

\newcommand{\Prog}[3]{{\sl Prog. Theor. Phys.} {\bf #1}  (#2) {#3}}
\newcommand{\IJMPA}[3]{{\sl Int. J. of Mod. Phys. A} {\bf #1}  {(#2)} {#3}}
\newcommand{\IJMPD}[3]{{\sl Int. J. of Mod. Phys. D} {\bf #1}  {(#2)} {#3}}
\newcommand{\GRG}[3]{{\sl Gen. Rel. Grav.} {\bf #1}  {(#2)} {#3}}




\end{document}